\documentclass[12pt,preprint]{aastex}
\usepackage{color}
\shorttitle{Delayed Onset of High-Energy Emissions in GRBs}
\shortauthors{Asano \& M\'esz\'aros}

\begin{document}

\title{
Delayed Onset of High-Energy Emissions in\\
Leptonic and Hadronic Models of Gamma-Ray Bursts
}
\author{\scshape Katsuaki Asano\altaffilmark{1},
and
Peter M\'esz\'aros\altaffilmark{2}}
\email{asano@phys.titech.ac.jp, nnp@astro.psu.edu}

\altaffiltext{1}{Interactive Research Center of Science, 
Tokyo Institute of Technology, 2-12-1 Ookayama, Meguro-ku, Tokyo 152-8550, Japan}
\altaffiltext{2}{Department of Astronomy \& Astrophysics; Department of Physics;
Center for Particle \& Gravitational Astrophysics;
Pennsylvania State University, University Park, PA 16802}

\date{Submitted; accepted}

\begin{abstract}

The temporal--spectral evolution of the prompt emission of gamma-ray bursts (GRBs)
is simulated numerically for both leptonic and hadronic models. 
For weak enough magnetic fields, leptonic models can reproduce the few 
seconds delay of the onset of GeV photon emission observed by {\it Fermi}-LAT, 
due to the slow growth of the target photon field for inverse Compton scattering. 
For stronger magnetic fields, the GeV delay can be explained 
with hadronic models, due to the long acceleration timescale of protons and 
the continuous photopion production after the end of the particle injection. 
While the FWHMs of the MeV and GeV lightcurves are almost the same in one-zone 
leptonic models, the FWHM of the 1--30 GeV lightcurves in hadronic models are 
significantly wider than those of the 0.1--1 MeV lightcurves. The amount of the GeV 
delay depends on the importance of the Klein--Nishina effect 
in both the leptonic and hadronic models.
In our examples of hadronic models the energies of
the escaped neutrons are comparable to the gamma-ray energy,
although their contribution to the ultra high-energy cosmic rays
is still subdominant.
The resulting neutrino spectra 
are hard enough to avoid the flux limit constraint from IceCube.  The delay 
of the neutrino emission onset is up to several times longer than the 
corresponding delay of the GeV photon emission onset.
The quantitative differences in the lightcurves for various models may be 
further tested with future atmospheric Cherenkov telescopes whose effective 
area is larger than that of {\it Fermi}-LAT, such as CTA.

\end{abstract}


\maketitle

\section{Introduction}
\label{sec:intro}

The nature of the prompt emission mechanism of gamma-ray bursts (GRBs) is 
still controversial, and remains a challenging problem in high-energy astrophysics.
Most of the GRB spectra peak around the 0.1--1
MeV range, and on the whole are
well fitted by the Band function \citep{ban93}. In the standard picture,
this component is explained by synchrotron emission from accelerated
electrons in a generic  dissipation region \citep{pir05,mes06} outside the
photosphere but inside the external shock radius. However, alternative models 
have been proposed such as photospheric emission \citep[see e.g.][]{rees05,
giannios07,bel10,ryd10}.
The recent detection of GeV photons with {\it Fermi}-LAT has opened up the
possibility of constraining such models. In some objects {\it Fermi} has also 
found in the GeV energy range additional spectral components 
\citep{510,510b,902B,926A}.
Moreover, a general feature is that the onset of the GeV emission tends to 
be delayed relative to the onset of the main MeV emission \citep[][and others 
including the above references]{916C}.
The typical delay timescales for the long {\it Fermi}-LAT GRBs are
3-5 s as seen in GRB 080916C, GRB 090902B and GRB 090926A \citep{mes12}.
The simplest model to explain this would be inverse Compton emission from the 
same internal dissipation region as the MeV \citep[][hereafter AM11]{cor10,asa11}.
A hadronic origin for the GeV emission have also been proposed \citep{asa09a,asa10,raz10},
although such models require a much larger energy in accelerated protons
than in the emitted gamma-ray energy.
Alternatively, if the Lorentz factor of the external shock propagating in the 
interstellar medium is as high as 1000, the delayed onset of the GeV emissions 
may be attributed \citep{ghi10,kum10} to an early onset of a forward shock 
synchrotron afterglow.
Other mechanisms to reproduce the GeV delay are proposed
by several authors \citep[see e.g.][]{tom09,iok10},
but we do not arrive at an established interpretation yet.

The photon statistics above the GeV range provided by {\it Fermi} are not
sufficient to distinguish between the internal or external origin of the
high-energy emission. However, it is expected that future multi-GeV observations 
with atmospheric Cherenkov telescope arrays such as CTA \citep{gil12,kak12,ino12}
will drastically improve the data quality, owing to their large effective area.
Lightcurves and spectral evolution determinations expected from such telescopes
should provide critical information on the GRB physics, currently being
pioneered by {\it Fermi}. To discriminate between the emission models, detailed
temporal--spectral evolution studies for various situations \citep[e.g.][]{pee05,
pee08,bel08, vur09,bos09,dai11} are needed.

In this paper, we concentrate on the internal dissipation regions
as possible models for explaining the delayed onset of the GeV emission.
In AM11, we developed a time-dependent code that can follow the evolution
of the particle energy distributions in a relativistically expanding shell
with Lorentz factor $\Gamma$ and initial radius $R_0$ from the central engine.
In this previous AM11 calculation we included only leptonic processes. However,
GRBs have been considered as possible sources of ultra high-energy cosmic rays
\citep[UHECRs,][]{wax95,vie95}, and the electromagnetic cascades initiated by 
photopion production from accelerated protons \citep[hadronic cascade,][]{boe98,gup07} 
result in both neutrinos and very high energy photons. The latter, potentially,
could reproduce the delayed onset of the GeV emission observed, because of the 
long proton-acceleration timescale and/or the continuous pion production even 
after the end of particle injection. Also, the flat spectrum of the keV--GeV 
power-law component seen in GRB 090902B \citep{902B} is compatible with a
hadronic cascade model \citep{asa10}. Thus, in order to explore the time
evolution of hadronic models, in this paper we incorporate the full details
of the hadronic processes into the time-dependent code from AM11.
With this expanded code, we are able to calculate the time evolution of
both the usual leptonic model, as well as the time evolution of the radiation 
from hadronic cascade models.

The observed lightcurves can show several pulses, and the origin of
the time variability timescale is unknown. One possibility for this 
is variability in the central engine ejection, in which case  the
observed pulse widths are limited by the usual formula $\sim R_0/c \Gamma^2$,
and there could be multiple MeV-photon sources before the GeV onset.
In this case, the GeV-delay is due to a superposition of the effects from various
shell sources with different  model parameters, the first pulse tending to 
have a soft spectrum in this scenario. Alternatively, the variability might
be due other effects, such as hydrodynamical turbulence in the shell 
or behind the shocks \citep{kum09,zha09,miz10,ino11}, with various
unconstrained parameters. To avoid such model dependent uncertainties,
in this paper we restrict ourselves to the study of a single primary 
pulse produced in a one-zone model, our code producing a smooth lightcurve, 
which can be considered as an envelope which averages a possible underlying 
variability with some smoothing. What we test is whether a single pulse in
such a one-zone models can reproduce the observed GeV delays or not, with
either leptonic or hadronic radiation mechanisms.
The differences in the lightcurves for the various models explored in this paper
provide clues which would be useful for discriminating between them, using future 
observations from atmospheric Cherenkov telescopes.

In \S \ref{sec:model}, we go into some of the details of the new numerical code 
for calculating the spectral evolution. The basic model features and our 
numerical results are presented in \S \ref{sec:lep} for the leptonic
and in \S \ref{sec:had} for the hadronic models, respectively. A discussion 
and summary of these results is given \S \ref{sec:sum}.

\section{Numerical Methods}
\label{sec:model}

The numerical code in this paper is developed from the one-zone code in AM11.
With this code we can simulate the temporal evolution of photon emission.
The photon source in this code is a shell expanding with the Lorentz factor $\Gamma$
from an initial radius $R=R_0$.
This shell is responsible for one pulse, as part of a lightcurve which may 
have multiple pulses.
The calculation of the photon production (for details see AM11)
is carried out in the shell frame (hereafter, the quantities in this frame
are denoted with primed characters). 
In this paper we express particle kinetic energies of electrons, protons, and photons
as $\varepsilon_{\rm e}=(\gamma_{\rm e}-1) m_{\rm e} c^2$,
$\varepsilon_{\rm p}=(\gamma_{\rm p}-1) m_{\rm p} c^2$ and
$\varepsilon$, respectively.

The code used in AM11 can simulate the injection and cooling for electrons/positrons,
and production, absorption and escape for photons with physical processes of
(1) synchrotron, (2) Thomson or inverse Compton (IC) scattering
(including the Klein--Nishina regime),
(3) synchrotron self-absorption (SSA),
(4) $\gamma \gamma$ pair production, (5) adiabatic cooling.
In the present paper, a new feature is the incorporation of the
hadronic processes (based on the simulation methods in the series of GRB 
studies of \cite{asa05,asa06,asa07,asa09b}) into the code of AM11.
For GRB prompt emission,
the most important process in hadronic cascade
is photopion production.
The timescale of photopion production in the one-zone
approximation is written as
\begin{eqnarray}
t^{\prime -1}_{{\rm p}\gamma}=\frac{c}{2} \int d \varepsilon'
\int_{-1}^1 d \mu'
(1-\mu')
n'_\gamma (\varepsilon') \sigma_{{\rm p}\gamma}
K_{{\rm p}\gamma},
\label{timescl}
\end{eqnarray}
where $n_\gamma (\varepsilon)$ is the energy distribution of the photon density,
$\mu$ is the cosine of the photon
incident angle, and $K_{{\rm p}\gamma}$
is the inelasticity.
We adopt experimental results for the
cross sections $\sigma_{{\rm p}\gamma}$ for
$p \gamma \to n \pi^+$,
$p \pi^0$, $n \pi^+ \pi^0$,
and $p \pi^+ \pi^-$ for
$\varepsilon'' \leq 2$ GeV \citep[see][for details]{asa06},
where $\varepsilon''$ is the photon energy in the proton
rest frame.
Multi-pion production due to high-energy gamma-rays
far above the $\Delta$-resonance ($\varepsilon'' \sim 300$ MeV) is not important
in our simulations \citep{mur06}, because the standard synchrotron model
generates a sufficiently soft photon spectrum with the typical index of $-1.5$.
For the pion production by $n \gamma$,
we adopt the same cross sections as $p \gamma$.
The inelasticity is approximated by a conventional
method as
\begin{eqnarray}
K_{{\rm p}\gamma}=\frac{1}{2} \left( 1-\frac{m_{\rm p}^2-m_\pi^2}{s} \right),
\end{eqnarray}
where $s$ is the invariant square
of the total four-momentum of the $p \gamma$
($n \gamma$) system,
and $m_{\rm p}$ is the proton mass.
For the double-pion production, we approximate
the inelasticity by replacing $m_\pi$ with
$2 m_\pi$.
Adopting the same Monte Carlo method as \citet{asa05}
or \citet{asa06}, we estimate energy loss of protons/neutrons
and inject pions every time step
following equation (\ref{timescl}).

For reference, we rewrite equation (\ref{timescl}) for a simple
case, in which a source with the bulk Lorentz factor $\Gamma$
emits photons at a radius $R$ with a power-law spectrum
$n'_\gamma (\varepsilon') \propto \varepsilon^{\prime \alpha}$ for
$\varepsilon' < \varepsilon'_{\rm peak}$
and luminosity $L_{\rm L}$ (integrated below $\varepsilon_{\rm peak}$).
Here, $\varepsilon_{\rm peak}$ corresponds to the spectral peak
energy of the Band function.
The result becomes
\begin{eqnarray}
t^{\prime -1}_{{\rm p}\gamma}= \frac{(2+\alpha) L_{\rm L}}
{8 \pi R^2 \Gamma^2}
\frac{2^{1-\alpha}}{1-\alpha}
\varepsilon_{\rm peak}^{\prime -(\alpha+2)}
\gamma^{\prime -(\alpha+1)}_{\rm p}
\int d \varepsilon'' K_{{\rm p}\gamma} \sigma_{{\rm p}\gamma} 
\varepsilon^{\prime \prime \alpha} \label{eqAtpg'},
\end{eqnarray}
for
\begin{eqnarray}
\gamma'_{\rm p} \gtrsim \gamma'_{\rm p,th}
\equiv \frac{300 \mbox{MeV}}{\varepsilon'_{\rm peak}}.
\end{eqnarray}
If we adopt the observationally typical index $\alpha=-1$,
the timescale does not depend on the energy of proton.
As is well known, the photomeson production efficiency
$f_{{\rm p}\gamma} \equiv
t'_{\rm exp}/t'_{{\rm p}\gamma}$ is close to unity for fiducial parameter sets
\citep[$t'_{\rm exp}=R/c\Gamma$,][]{wax97,asa05,mur06}.
The theoretically expected value $\alpha=-1.5$ in the standard synchrotron model
leads to more efficient pion production for higher energy protons
as $f_{{\rm p}\gamma} \propto \gamma_{\rm p}^{0.5}$.
Additionally, the hadronic cascade may produce an even softer spectrum.
Our time-dependent simulation is a powerful tool to follow
the pion production rate as the photon spectrum evolves.

In terms of the Thomson optical depth $\tau_{\rm T}$, the $pp$ reaction 
efficiency is $f_{\rm pp} \sim 0.05 \tau_{\rm T}$ \citep{mur08}. Thus, 
we can neglect the $pp$-collisions in the optically thin cases in this paper.
However, since our code is planned to apply also to optically thick cases 
in the near future, we have mounted the $pp$ reaction on our code, as was done 
in \citet{mur12}.
Calculated with the numerical simulation kit Geant4 \citep{Ago+03},
we provide tables for the cross section, inelasticity and
pion-multiplicity for $pp$-collision.
The procedure to follow proton cooling and pion injection
is the same as that for $p \gamma$-collision.

The Bethe--Heitler (BH) pair production process ($p \gamma \to p e^+ e^-$)
is also taken into account.
The cooling and spectral pair injection rates are calculated
with the cross section and inelasticity from \citet{cho92}.

In our code, neutral pions promptly decay into two gamma-rays,
while the cooling of charged pions before their decaying is taken into account.
For charged particles, such as protons, pions and muons,
synchrotron and IC with the full Klein--Nishina cross section
are included as photon production and cooling processes.
The particle decays are simulated with the Monte Carlo method
adopting the lifetime of pions (muons) as $2.6 \times 10^{-8} \gamma_\pi$ s
($2.2 \times 10^{-6} \gamma_\mu$ s).
The energy fraction of muons at pion decay
$\pi^+(\pi^-) \to \mu^+ \nu_\mu (\mu^- \bar{\nu}_\mu)$
is approximated as $m_\mu/m_\pi \sim 0.76$,
and the rest of the energy
goes to a neutrino. 
On the other hand, we assume that the energy of a muon at its
decay $\mu^+ (\mu^-) \to e^+ \nu_{\rm e} \bar{\nu}_\mu
(e^- \bar{\nu}_{\rm e} \nu_\mu)$
        will be shared equally by a positron (electron), neutrino,
and antineutrino.
We neglect neutron decay, whose timescale is much longer than
all the timescales we consider in this paper.

We also take into account the effect of adiabatic cooling
for charged particles with the same method in AM11.
This effect is controlled by the expansion law
of the volume $V'=4 \pi R^2 W'$, where
$W$ is the shell width.
The volume expansion law affects the escape rate
for neutral particles.
As mentioned in AM11, 
the escape fraction of photons per unit time
is $c/2 W'$.
We consider several cases for the expansion law in our simulations.

Assuming the jet opening angle of $\theta_{\rm jet} = 10/\Gamma$,
the photon spectrum evolution for an observer is calculated
in the same manner in AM11.
The energy and escape time of photons coming from a surface
with an angle $\theta$
(measured from the central engine,
$\theta=0$ corresponds to the line of sight)
are transformed into energy- and time-bins in observer's frame as
$\varepsilon=\delta \varepsilon'/(1+z)$ ($\delta$ is the Doppler factor)
and
\begin{eqnarray}
t_{\rm obs}=(1+z) \left[ (1-\beta_{\rm sh} \cos{\theta}) \Gamma t'
+R_0 (1-\cos{\theta})/c \right],
\end{eqnarray}
respectively.
When $\cos \theta < \beta_{\rm sh} \equiv \sqrt{1-1/\Gamma^2}$,
photons are coming from the backside of the shell
with an extra ``time delay'' due to the shell thickness,
\begin{eqnarray}
\Delta t_{\rm ex}=(1+z)\frac{W' \Gamma}{c} \left[
\left( \beta_{\rm sh}^2 +1/\Gamma^2
\right) \cos{\theta}-\beta_{\rm sh} \right],
\end{eqnarray}
relative to the emission from the fore side.

The photon absorption due to
extra galactic background light (EBL) is estimated by accumulating
the optical depth between the source and observer
with the EBL model of \citet{kne04} (best-fit).

\section{Leptonic Models}
\label{sec:lep}

\subsection{Delayed Inverse Compton}

Inverse Compton emission, which may dominate the GeV energy range,
needs low-energy seed photons.
In the synchrotron self-Compton model (SSC),
the seed photons are synchrotron photons emitted from
power-law distributed electrons.
The growth timescale of the synchrotron photon field
may explain the delay timescale of the GeV onset \citep{bos09}.
This possibility was tested by time-dependent simulations in AM11.
They showed that a very low magnetic field ($\epsilon_B/\epsilon_{\rm e} \sim 10^{-3}$
and $R_0=6 \times 10^{15}$ cm in their example) is required to reproduce the GeV delay.
Here, we write the magnetic field
\begin{eqnarray}
B'&=&\sqrt{\frac{2 \epsilon_B L_{\rm iso}}{\epsilon_{\rm e} c R^2 \Gamma^2}} \\
&\simeq& 320 \left( \frac{\epsilon_B/\epsilon_{\rm e}}{10^{-3}} \right)^{1/2}
\left( \frac{\Gamma}{800} \right)^{-1}
\left( \frac{R}{10^{15}\mbox{cm}} \right)^{-1}
\left( \frac{L_{\rm iso}}{10^{54}\mbox{erg}~\mbox{s}^{-1}} \right)^{1/2} \mbox{G}.
\label{eqB}
\end{eqnarray}
The standard synchrotron model attributes the spectral peak energy
$\varepsilon_{\rm peak}$ to the typical photon energy emitted
from the lowest-energy electrons at injection.
The physical mechanism that determines the minimum-Lorentz factor
$\gamma_{\rm e,min}$ of those electrons is still under discussion.
In a low magnetic field as given in eq. (\ref{eqB}),
to emit synchrotron photons
in MeV range,
a very high minimum-Lorentz factor ($\gg m_{\rm p}/m_{\rm e}$) is required as
\begin{eqnarray}
\gamma'_{\rm e,min} &\sim& \sqrt{\frac{\varepsilon_{\rm peak} m_{\rm e} c}
{\hbar e B' \Gamma}} \\
&\simeq& 1.8 \times 10^4
\left( \frac{\varepsilon_{\rm peak}}{\mbox{MeV}} \right)^{1/2}
\left( \frac{\epsilon_B/\epsilon_{\rm e}}{10^{-3}} \right)^{-1/4}
\left( \frac{R}{10^{15}\mbox{cm}} \right)^{1/2}
\left( \frac{L_{\rm iso}}{10^{54}\mbox{erg}~\mbox{s}^{-1}} \right)^{-1/4},
\end{eqnarray}
where we have omitted the cosmological redshift factor.
In this case the typical photon energy in the electron rest frame becomes
\begin{eqnarray}
\varepsilon''_{\rm peak}&\sim&\gamma'_{\rm e,min} \varepsilon'_{\rm peak}= \gamma'_{\rm e,min}
\varepsilon_{\rm peak}/\Gamma \\
&\simeq& 23
\left( \frac{\varepsilon_{\rm peak}}{\mbox{MeV}} \right)^{3/2}
\left( \frac{\epsilon_B/\epsilon_{\rm e}}{10^{-3}} \right)^{-1/4}
\left( \frac{\Gamma}{800} \right)^{-1}
\left( \frac{R}{10^{15}\mbox{cm}} \right)^{1/2}
\left( \frac{L_{\rm iso}}{10^{54}\mbox{erg}~\mbox{s}^{-1}} \right)^{-1/4} \mbox{MeV},
\end{eqnarray}
which largely exceeds $m_{\rm e} c^2$, so the Klein--Nishina effect
is crucial to emit IC photons.
The IC photon production is an inefficient process so that
the GeV component shows a slow growth compared to synchrotron.
Soft photons ($\varepsilon \ll \varepsilon_{\rm peak}$)
that can efficiently interact with electrons of $\gamma_{\rm e,min}$
are produced by electrons that have already undergone energy losses
and have $\gamma_{\rm e} < \gamma_{\rm e,min}$.
However, as soft photons increases, the newly injected electrons cool
via mainly IC emission owing to a very low $\epsilon_B/\epsilon_{\rm e}$.
Therefore, the synchrotron component starts decreasing while
the IC component grows.
This enhances the difference in the peak times of the MeV and GeV lightcurves.

For the purposes of comparison with the leptonic calculations in 
the present paper, we summarize here some of the leptonic results in AM11.
The lightcurves in Figure 12 in AM11 show
a GeV delay due to the mechanism we mentioned above.
However, the difference in the peak times is within a
pulse timescale, $\sim R/c \Gamma^2$
(the difference in the peak times is roughly half of this timescale
in the example in AM11),
and the GeV and 100 keV lightcurves converge in the late stages.
The FWHMs of the two lightcurves are almost the same.
The example in AM11 adopted an ideal set of parameters for the GeV delay,
and it seems difficult to produce even larger delays by simple SSC models.
We did not find significant GeV delays for more conservative 
SSC parameter sets in our simulations in AM11.

An external inverse Compton (EIC) model \citep{bel05,tom09,tom10,li10,mur11}
was also tested in AM11.
The MeV photons are emitted from a smaller radius,
and they are upscattered by non-thermal electrons
in an outer region, where some dissipation mechanism, such as
internal shocks, inject those electrons.
In this model, we require two different origins
for the MeV and GeV components. A GeV delay larger than the 
typical pulse timescale ($R/c \Gamma^2$) appears naturally, although 
the model requires a large number of accelerated electrons,
which leads to an energy-budget problem.
If the numbers of electrons and protons are the same,
most of the energy carried by the protons remains unreleased.
Some GRBs, in which the main MeV component is hard to explain by 
the usual synchrotron model, like GRB 090902B \citep{902B,ryd10,zha11},
are interesting applications for this model.
In the frame of the outer dissipation region,
the photon field coming from an inner region is highly beamed
along the radial direction.
The $e \gamma$-scattering probability becomes highest for
head-on collision so that the scattered IC-photons
tend to propagate with a large angle to the radial axis.
As a result, the intensity of the scattered photons
becomes anisotropic. In this case, the emission from 
off-axis regions ($\theta \gtrsim 1/\Gamma$) contributes considerably.
AM11 showed that the GeV lightcurve has a long tail
(see Figure 15 in AM11), which is a characteristic feature of this model.

\subsection{Opacity Evolution in Leptonic Model}
\label{sec:opa}

As a shell expands, the photon density decreases so that
the opacity against $\gamma \gamma$ absorption also decreases
with time.
This effect may cause a gradual increase of the $\gamma \gamma$ cut-off energy
\citep{gra08}.
In this section, we test the evolution of the $\gamma \gamma$-opacity
as a possible mechanism of the GeV delay.
When the electron injection timescale is comparable to
or shorter than the shell expansion timescale $t'_{\rm exp}=R_0/\Gamma c$,
the opacity evolution tends to cause a negative GeV delay:
the GeV emission is terminated earlier than the MeV emission
as shown in Figure 3 or 8 in AM11.
As the electron injection builds up the photon density,
$\gamma \gamma$-optical depth increases, and
GeV photons begins to be absorbed.
It was hard to observe the effect of the
opacity decrease in our simulations with an injection
timescale of $R_0/\Gamma c$.
Therefore, here we adopt a very long electron injection
to see the opacity decay as the shell expands.

We consider a shell expanding from an initial radius $R_0=10^{15}$ cm
with a bulk Lorentz factor $\Gamma=500$, and optimize parameters
to observe the opacity effects as below.
If we take a standard assumption for the shell width as $W'=R_0/\Gamma$,
the photon escape timescale and shell expansion timescale
become comparable.
In such a case, the opacity evolution may be softened by
residual photons generated earlier but before escape.
So we assume a thin shell $W'=R_0/10 \Gamma$.
The electron injection continues until $R=10^{16}$ cm
($t'_{\rm inj}=9 t'_{\rm exp}$) with a constant rate, and its isotropic-equivalent energy
for this pulse is $E_{\rm e,pls}=2 \times 10^{54}$ erg.
The injection spectrum is assumed as a cut-off power-law shape,
$\dot{N}'_{\rm e,inj}(\varepsilon'_{\rm e})
\propto \varepsilon^{\prime -p_{\rm e}}_{\rm e} \exp
(-\varepsilon'_{\rm e}/\varepsilon'_{\rm e,max})$
for $\varepsilon'_{\rm e}>\varepsilon'_{\rm e,min}$,
where $\varepsilon'_{\rm e,max}$ is determined
by equating the cooling timescale $t'_{\rm cool}$
(synchrotron, IC, and SSA are taken into account)
and the acceleration timescale
\begin{eqnarray}
t'_{\rm acc}=\xi \frac{\varepsilon'_{\rm e}}{ceB'}.
\label{acctim}
\end{eqnarray}
In this section we assume the electron index $p_{\rm e}=2.5$,
and ``Bohm limit'' acceleration as $\xi=1$.
In order to avoid contamination of other evolutionary effects,
the magnetic field $B'=3.9 \times 10^4$ G, shell width,
and minimum injection energy of electrons $\varepsilon'_{\rm e,min}=2$ GeV
are assumed to be constant during the electron injection.
However, the ratio $\epsilon_B/\epsilon_{\rm e}$, which may be approximated
as
\begin{eqnarray}
\frac{\epsilon_B}{\epsilon_{\rm e}}
\simeq \frac{B'^2}{8 \pi} \left( \frac{1}{4 \pi R^2 W'}
\frac{E_{\rm e,pls}}{\Gamma}
\frac{t'_{\rm esc}}{t'_{\rm inj}}
\right)^{-1}
=\frac{9 R_0 R^2 \Gamma B'^2}{2 E_{\rm e,pls}},
\end{eqnarray}
evolves from $0.04$ to $4$ during the electron injection.

\begin{figure}[htb!]
\centering
\epsscale{1.0}
\plotone{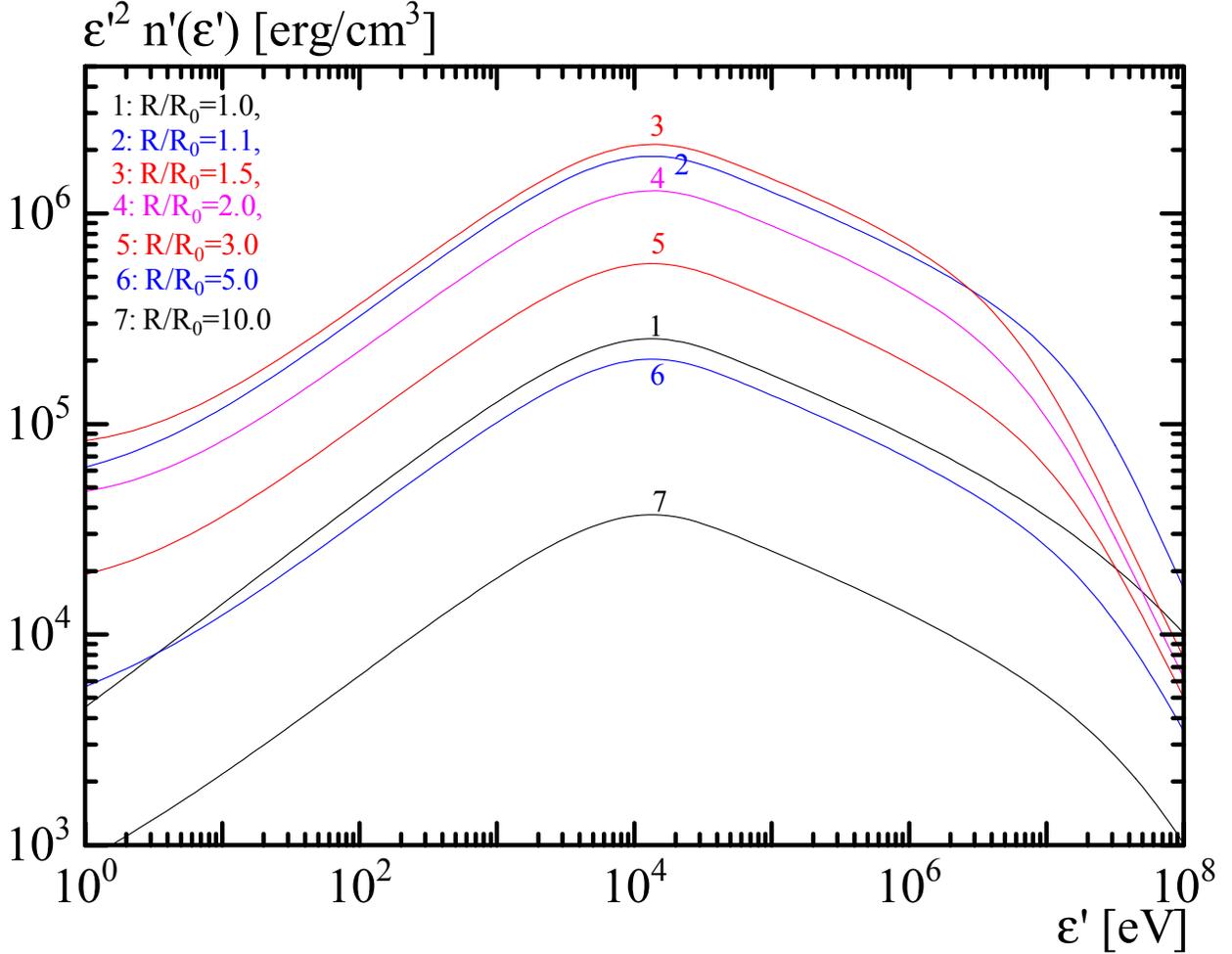}
\caption{Opacity evolution leptonic model: spectral evolution of photon 
density in the shell frame with the expanding radius $R$
(see the text in \S \ref{sec:opa}).
Initially the photon density increases as electrons are injected
($R<1.5R_0$), then the shell expansion and photon escape make
the density begin to decrease.
In this stage the break energy $\varepsilon'_{\rm br} \sim 10^7$ eV 
due to $\gamma \gamma$-absorption gradually grows as the shell expands.
\label{fig:1}}
\end{figure}

Figure \ref{fig:1} shows the spectral evolution in the shell frame.
The spectral peak energy $\varepsilon'_{\rm peak} \sim 10$ keV
arising from the parameter set chosen here is somewhat higher than the 
usual values.  This is because we are considering a special case similar
to GRB 080916C, which showed a delayed GeV onset with a source-frame peak 
energy of $(1+z) \varepsilon_{\rm peak,obs} \sim 6$ MeV \citep{916C}.
In a strict sense, $R=R_0$ in Figure \ref{fig:1} means that
a short time $t'_{\rm exp}/100$ has passed after the electron injection started.
At this point, the spectrum is not in a steady state for photon production,
escape, and $\gamma \gamma$ absorption.
During the initial phase for $R \leq 1.5 R_0$,
the opacity still increases with the photon density.
At $R=1.5 R_0$ the high-energy portion of the spectrum can be approximated
by a broken power-law with a photon index $\beta \sim -2.3$ below the break
energy $\varepsilon'_{\rm br}$ and $-3.3$ above that, 
rather than an exponential cut-off.
This break is similar to the prediction by \citet{gra08},
in which a thin slab is considered as a photon source.
However, the mechanism for yielding a broken power-law
is different for our one-zone model with a finite shell width.
In a quasi steady state, the fraction of photons, whose
annihilation timescale $t'_{\rm ann}$ is shorter
than the escape timescale $t'_{\rm esc} \sim W'/c$,
may be written as $t'_{\rm ann}/t'_{\rm esc}$.
Then, the spectral shape may be $n'_0(\varepsilon') t'_{\rm ann}/t'_{\rm esc}$,
where $n'_0(\varepsilon') \propto \varepsilon^{\prime \beta}$
is the intrinsic spectrum,
whose density is determined by the balance between
continuous photon production and escape/volume-expansion without the absorption effect.
The annihilation timescale is $t'_{\rm ann} \propto \varepsilon^{\prime \beta+1}$
\citep{lit01,asa03},
while $t'_{\rm esc}$ is independent of the photon energy.
Therefore, $n'(\varepsilon') \propto \varepsilon'^{2 \beta+1}$
above the spectral break at $\varepsilon'_{\rm br} \sim 10^7$ eV.
The value $\beta \sim -2.3$ implies that
the index should be $-3.6$.
We should note that photons above $\varepsilon'>10^7$ eV interact with
photons below $\varepsilon'_{\rm peak}$.
In our example in Figure \ref{fig:1},
the index of the photon distribution for
one order of magnitude just below $\varepsilon'_{\rm peak}$
is $\alpha \sim -1.3$.
In such cases, the spectrum should be $n'(\varepsilon') \propto
\varepsilon'^{\beta+\alpha+1} \sim \varepsilon'^{-3.0}$.
Our result seems consistent with those estimates.

In the later stage ($R \geq 2 R_0$), the $\gamma \gamma$-attenuation gradually
becomes inefficient, and this causes the photon spectral break energy to increase with time.
In this stage the photon distribution may be in the quasi steady state.
Therefore, the spectra for $R=R_0$ (photon-density growing stage)
and $R=5 R_0$ are different
in spite of their similar photon densities.
The deviation from a simple power-law seen around 1 eV
is due to electron heating via SSA \citep[][AM11]{ghi88}.

\begin{figure}[htb!]
\centering
\epsscale{1.0}
\plotone{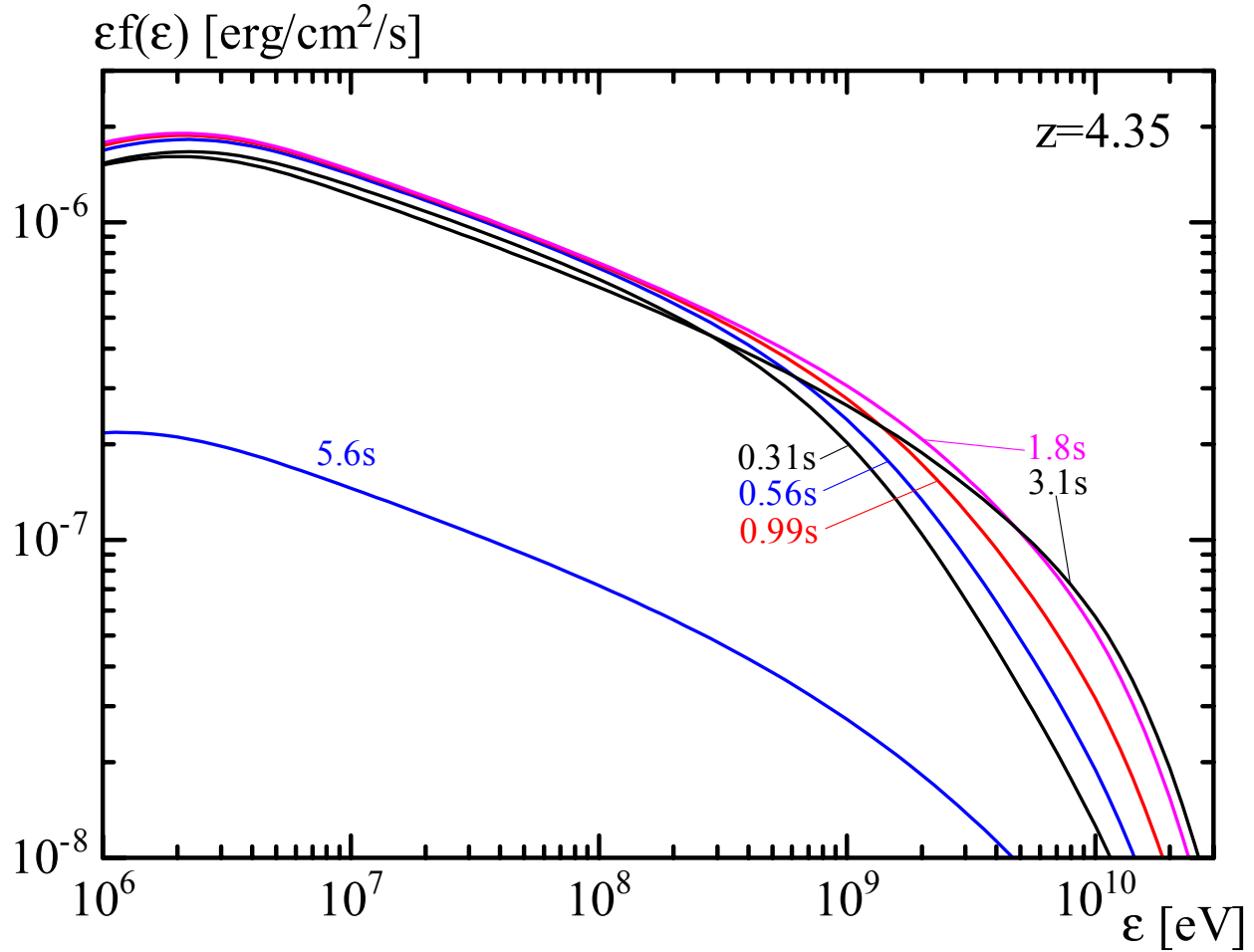}
\caption{Opacity evolution leptonic model: flux evolution for an observer 
with $z=4.35$
(see the text in \S \ref{sec:opa}).
\label{fig:2}}
\end{figure}

Assuming a source redshift $z=4.35$ such as that of as GRB 080916C, we show
the evolution of the spectrum for an observer in Figure \ref{fig:2}.
Given a luminosity $L_{\rm H}$ above $\varepsilon_{\rm peak}$,
in a case of the usual short electron-injection
($t'_{\rm inj} \leq R_0/\Gamma c$),
an analytical estimates \citep{lit01,asa03}
gives us the $\gamma \gamma$ cut-off/break energy as
\begin{eqnarray}
\varepsilon_{\gamma \gamma}=
\frac{m_{\rm e}^2 c^4}{\varepsilon_{\rm peak}} \Gamma^{2 \frac{\beta-1}{\beta+1}}
\left[ \frac{\sigma_{\rm T} L_{\rm H}}{16 \pi \varepsilon_{\rm peak} c^2 \delta t}
F(\beta)
\right]^{\frac{1}{\beta+1}} \propto \delta t^{-\frac{1}{(\beta+1)}},
\end{eqnarray}
where the dimensionless function $F(\beta) \sim 10^{-2}$,
and $\delta t=R/\Gamma^2 c$ is the variability timescale.
For our continuous injection case,
that may be interpreted as $\varepsilon_{\gamma \gamma} \propto t_{\rm obs}^{0.8}$
with $\beta=-2.3$.
By comparing spectra for $t_{\rm obs}=0.31$ s and $3.1$ s,
we can see that the spectral behavior
is close to this analytical approximation;
the break energy grows by a factor of six or more.

\begin{figure}[htb!]
\centering
\epsscale{1.0}
\plotone{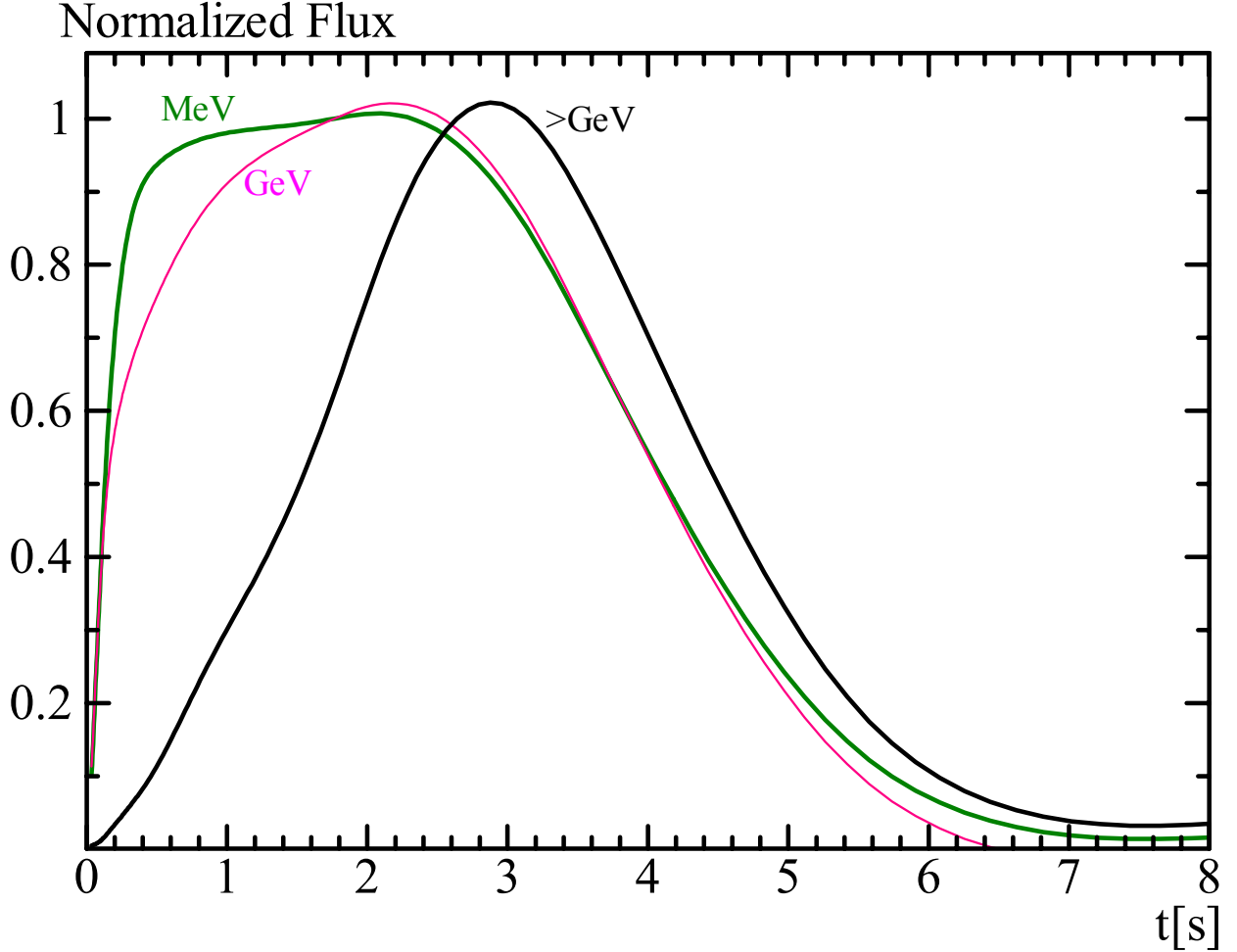}
\caption{Opacity evolution leptonic model lightcurve (see the text in \S \ref{sec:opa}).
The source redshift is assumed to be $4.35$.
The lines labeled with ``GeV'' or ``MeV'' are plotted based on
the spectral flux, while the line labeled with ``$>$GeV''
is a curve based on the photon number integrated above GeV.
\label{fig:3}}
\end{figure}

However, the spectral shape does not show a sharp cut-off
or break.
The deviation from the extrapolation of the low-energy spectral shape
is gradual.
As a result, the spectral flux at GeV (see ``GeV'' in Figure \ref{fig:3})
shows a similar evolution to the flux at MeV.
The MeV lightcurve has a plateau reflecting the constant
electron injection rate.
The peak time of the GeV lightcurve delays relative to the
MeV arising time, but the GeV flux changes slowly during
the MeV-plateau phase.
However, for low photon statistics, the lightcurve
would be obtained by accumulating photon counts above GeV.
As shown in Figure \ref{fig:3}, the lightcurve based on
the accumulated photon counts shows a distinct delay
relative to the MeV lightcurve.

We have considered an artificial setup to see a GeV delay
due to the opacity evolution.
Even in this extreme case, the spectral evolution is gradual, and it seems 
different from the observed ``sudden'' onset of GeV emission.

\section{Hadronic Models}
\label{sec:had}

The electromagnetic cascade triggered by photopion production
can make GeV extra components as shown in our series of GRB studies
\citep{asa07,asa09a,asa09b,asa10}.
The electron cooling timescale is much shorter than
the timescales of proton acceleration and photopion production.
Those timescales and
continuous photomeson production after the end of electron injection
may cause a delayed onset of the GeV emission.

In this section, we inject electrons and protons with the same timescale
$t'_{\rm inj}=R_0/\Gamma c$.
The injection spectra have the same shape as that for electrons
in \S \ref{sec:opa}.
The proton injection index is assumed to be $p_{\rm p}=2.0$,
which is appropriate to make GRBs contribute to UHECRs
\citep{wax98},
while that for electrons is taken to be $p_{\rm e}=3.0$
to reproduce the typical photon index $\beta \sim -2.5$
measured by BATSE \citep{pre00,kan06}.
Although the injection indices for protons and electrons
are different, the energy scales we consider
are largely different for electrons and protons
\citep[see][for further discussion]{asa09b}.
We adopt the parameter $\xi=1$ for both electrons and protons
(see eq. (\ref{acctim})).
The maximum energy of protons is determined by equating the cooling
and acceleration timescales or by the condition that the
Larmor radius should be shorter than the shell width.
We take into account proton synchrotron and photopion production
to estimate the cooling timescale.
While the proton maximum energy at injection is controlled
by the time step for injection,
we phenomenologically carry out succeeding proton acceleration
following the acceleration timescale and index.
The minimum energy of the protons is fixed at 3 GeV.

We find that the shell width expansion enhances the GeV delay in our simulations.
Below, we show examples with an expanding width as $W'=R/\Gamma$.
During the particle injection, we neglect adiabatic cooling
and decay of the magnetic field.
After the particle injection, the magnetic field is
assumed to decay as $B' \propto R^{-2}$.

In this paper the MeV photon peak is assumed to be produced by
accelerated electrons injected in the emission region, which differs
from the previous assumption in \citet{asa09a,asa10}, where an ad-hoc 
Band-type photon component matching the observed spectra was postulated.
Therefore, differences from the preceding simulations arise not only 
because here we consider the time-dependence, but also because some of the 
basic input assumptions are different. One issue here is that the
synchrotron emission from the primarily accelerated electrons
yields a soft photon index $\alpha \sim -1.5$, which is difficult for
reproducing the observed spectral indices.
This is because the effective energy-distribution
for cooled electrons below $\gamma_{\rm e,min}$ becomes
$n_{\rm e}(\gamma_{\rm e}) \propto \gamma_{\rm e}^{-2}$,
when the synchrotron emission
is the dominant cooling process ($\dot{\gamma_{\rm e}} \propto \gamma_{\rm e}^2$).
In order to resolve this 
contradiction, another mechanism for the Band component is required,
such as a photosphere model or continuous particle acceleration
by turbulence \citep{asa09c,mur12}. However, since our emphasis here is more
on the GeV components, for simplicity we consider here the standard synchrotron 
model.

\subsection{ Hadronic ``Moderate'' Case}
\label{sec:mod}

Here we adopt a typical GRB luminosity for one pulse in a GRB,
as opposed to the high values deduced for LAT bursts.
The model parameters are $\Gamma=800$, $R_0=10^{14}$ cm,
$E_{\rm e,pls}=5.0 \times 10^{50}$ erg,
and an isotropic-equivalent energy of protons $E_{\rm p,pls}=6.1 \times 10^{51}$ erg
($\epsilon_{\rm p}/\epsilon_{\rm e} \simeq 12$)\footnote{An alternate notation in some
papers is $1/f_e\simeq \epsilon_{\rm p}/\epsilon_{\rm e}$} including the acceleration 
effect after injection.
The initial magnetic field $B'_0$ is set through the parameter ratio
at $R=2 R_0$ as
\begin{eqnarray}
\frac{\epsilon_B}{\epsilon_{\rm e}} \simeq \frac{4 R_0^3 B_0^{\prime 2}}
{E_{\rm e,pls}}=3.0,
\end{eqnarray}
which results in $B'_0=1.9 \times 10^4$ G.
The peak energy $\varepsilon_{\rm peak}$ is adjusted to be
$\sim 100$ keV by adopting $\varepsilon'_{\rm e,min}=380$ MeV.

\begin{figure}[htb!]
\centering
\epsscale{1.0}
\plotone{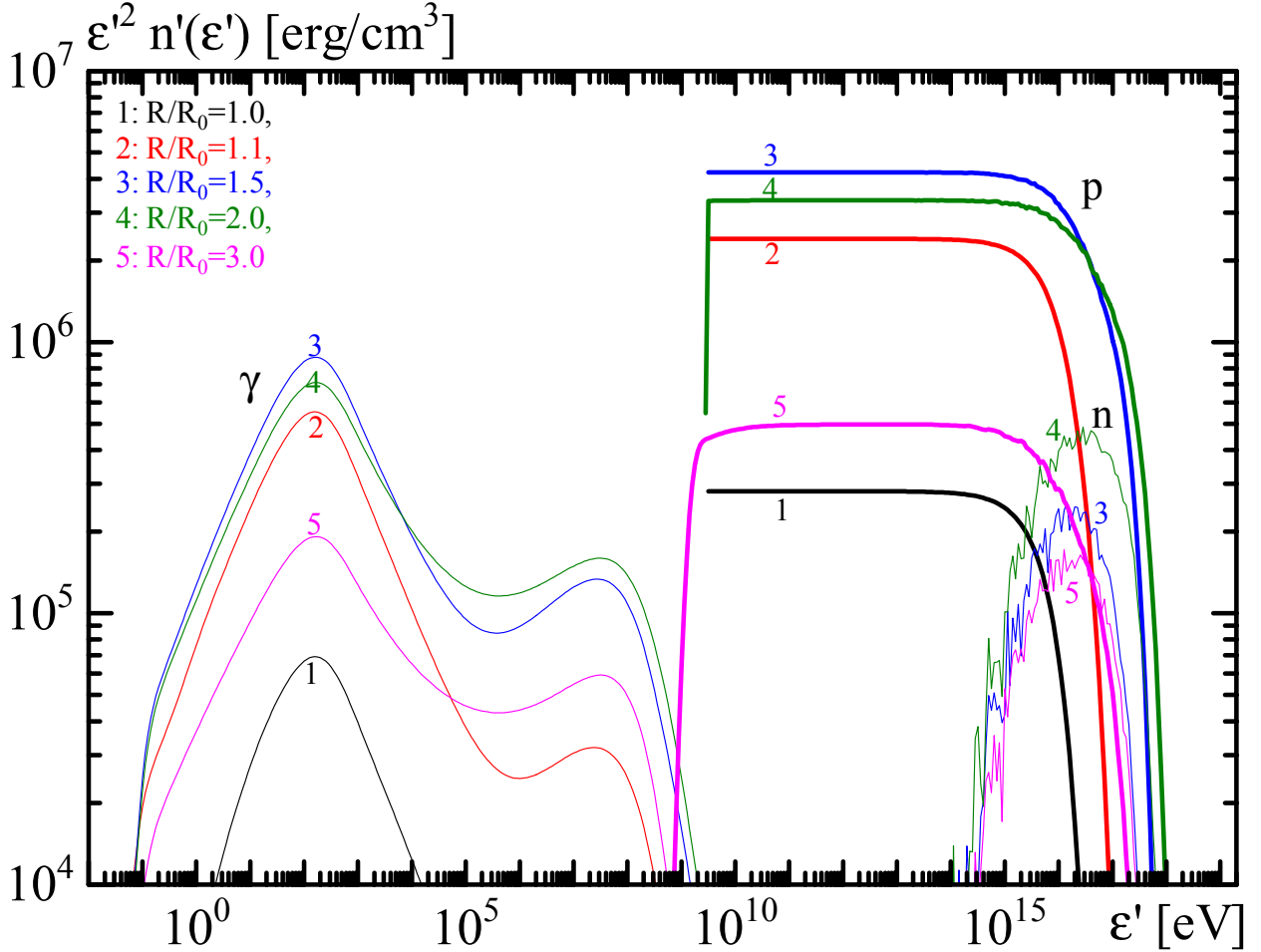}
\caption{Hadronic model, ``moderate" case: spectral evolution of the photons, 
protons, and neutrons in the shell frame with the expanding radius $R$
(see the text in \S \ref{sec:mod}).
As particle injection proceeds, the photon and proton densities build up.
Then, for $R>1.5 R_0$, the density starts to decrease due to the effects of
the shell expansion and photon escape.
The electromagnetic cascade triggered by photopion production gradually
enhances the height of the second peak ($\varepsilon' \sim 10^8$ eV) relative
to that of the main peak ($\varepsilon' \sim 10^2$ eV) in the photon spectrum.
\label{fig:4}}
\end{figure}

\begin{figure}[htb!]
\centering
\epsscale{1.0}
\plotone{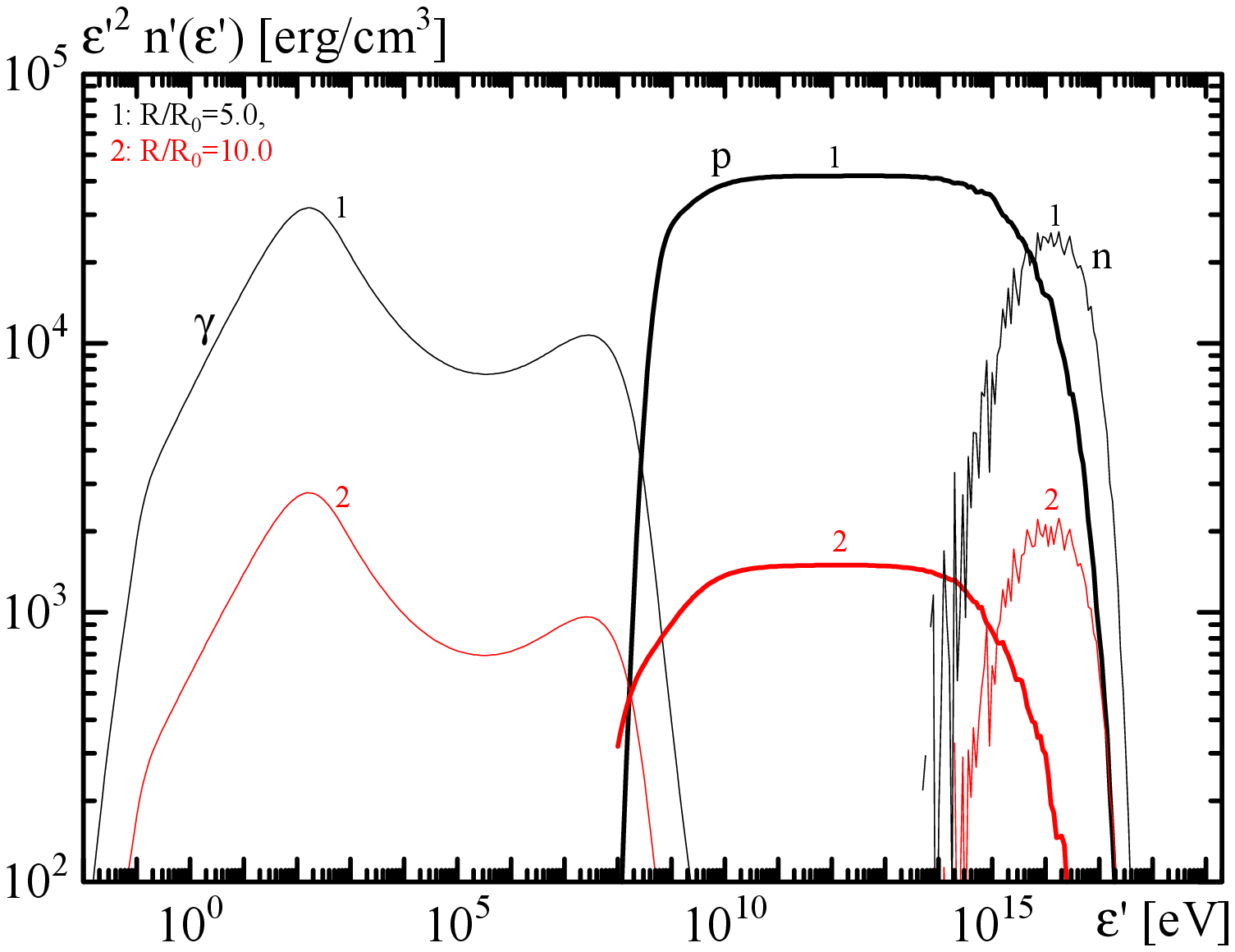}
\caption{ Hadronic model, ``moderate" case:
same as Figure \ref{fig:4} but for later stage.
Neutrons, which, being neutral, are not subject to adiabatic cooling, 
dominate the high energy region above $10^{15}$ eV.
\label{fig:5}}
\end{figure}

In Figures \ref{fig:4} and \ref{fig:5}, the main particle spectra are plotted in
the shell frame, omitting for clarity the spectra of the electron/positron, pion, 
muon, and neutrino components.
In the early phase ($R<2 R_0$), the proton density is increasing
by proton injection, and their maximum energy
is growing with the radius $R$ obeying the proton acceleration timescale.
Then, protons undergo adiabatic cooling, while neutrons keep their energies
until they escape or interact with photons.
The volume expansion ($V' \propto R^3$) reduces the particle number densities,
and particle escape also affects the densities of neutrons and photons.
The evolution of the neutron-to-proton ratio from $2 R_0$ to $3 R_0$
indicates that pion (and neutron as well) production continues
even after the end of particle injection.
Therefore, the electromagnetic cascade due to photopion process continues as well
for $t'>t'_{\rm inj}$.

In the photon spectra, we can see a $\gamma \gamma$ cut-off at $\sim 10^8$ eV
and an SSA (synchrotron self-absorption) signature at $\sim 0.1$ eV.
The second peak of the gamma-ray spectra at $\sim 10^8$ eV
is due both to SSC by the primarily accelerated
electrons and to synchrotron emission by secondary electrons/positrons.
The cascade due to photopion process raises the second peak relative
to the synchrotron-first peak even for $R>2 R_0$.
At $R=10 R_0$, the highest energy region ($>10^{16}$ eV) is dominated
by neutrons, because protons have been adiabatically cooled.

\begin{figure}[htb!]
\centering
\epsscale{1.0}
\plotone{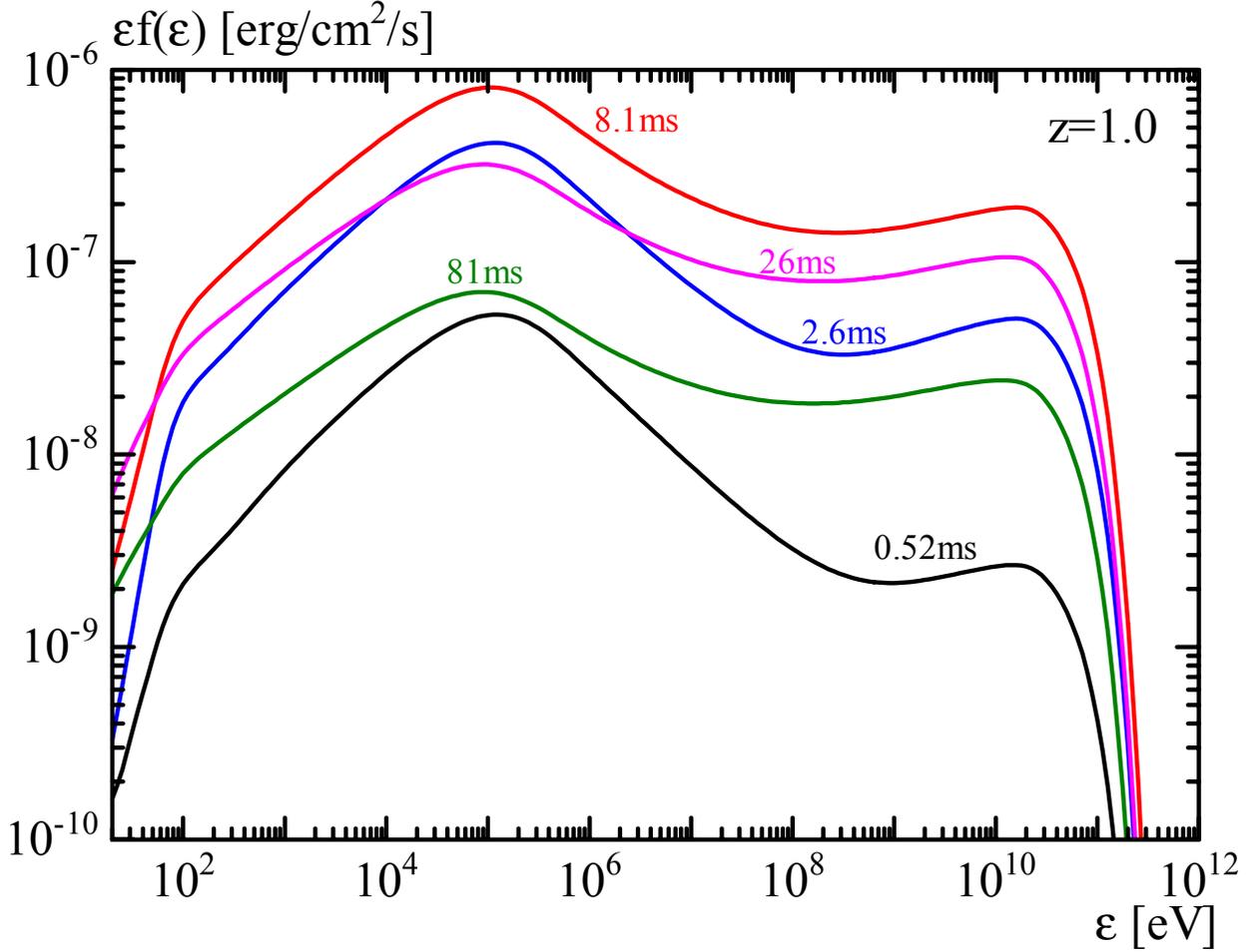}
\caption{Hadronic model, ``moderate" case: flux evolution
(see the text in \S \ref{sec:mod}) for an observer with $z=1$.
\label{fig:6}}
\end{figure}

Assuming $z=1.0$, we plot the spectral flux evolution of gamma-rays
for an observer in Figure \ref{fig:6}.
It is clearly seen that the spectral shape above $\varepsilon_{\rm peak}$
becomes harder with time,
and the fraction of the GeV--10 GeV component grows.
The softening in the lower-energy region ($\varepsilon < \varepsilon_{\rm peak}$)
is also caused by the hadronic cascade.
\begin{figure}[htb!]
\centering
\epsscale{1.0}
\plotone{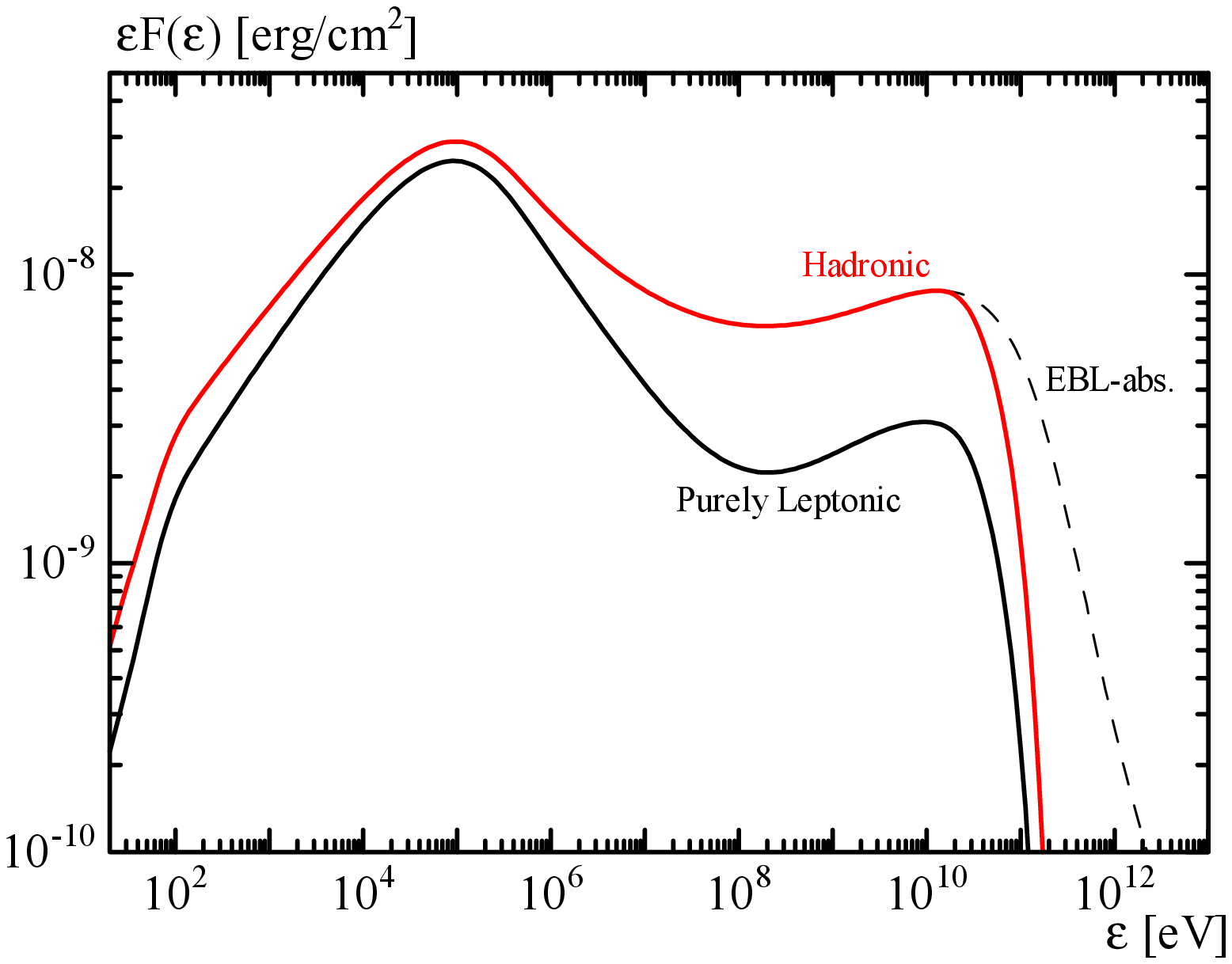}
\caption{Fluences for the hadronic model
and purely leptonic model ($\epsilon_{\rm p}=0$) with the ``moderate'' parameters
(see the text in \S \ref{sec:mod}) for an observer with $z=1$.
The dashed line is the spectrum without the absorption effect of EBL.
\label{fig:7}}
\end{figure}
The time-integrated spectrum is shown in Figure \ref{fig:7},
in which we compare this hadronic model with the same (leptonic) model 
without proton injection.
Both models show a second peak at $10^{10}$ eV, in the leptonic case this
being due to the SSC emission, while in the hadronic it is due to the cascade.
Thus, it is hard to distinguish these model just from their  spectral shape.
However, as shown in Figure \ref{fig:8}, the lightcurves for the hadronic model
(solid lines) may be distinguishable from those for the leptonic models 
(dashed lines). One sees that the 100 keV and 10 GeV lightcurves for the 
leptonic model (dashed lines) have almost the same shape,
and the GeV delay does not appear in this leptonic case.
In order to produce a GeV delay in these leptonic models, we may need a smaller 
magnetic field, such as $\epsilon_B/\epsilon_{\rm e}<10^{-3}$ or $B' \lesssim 100$ G,
as shown in AM11. 
On the other hand, for the ``moderate" hadronic model, the 10 GeV or 30 GeV 
lightcurves show a distinct (even if short) delay relative to the 100 keV lightcurve.
Even so, the peak time for the 10 (30) GeV-lightcurve is earlier
than the peak time for the neutrino lightcurve.
This may be because the SSC contribution from primary electrons
is not negligible in the 10 (30) GeV-lightcurve.

\begin{figure}[htb!]
\centering
\epsscale{1.0}
\plotone{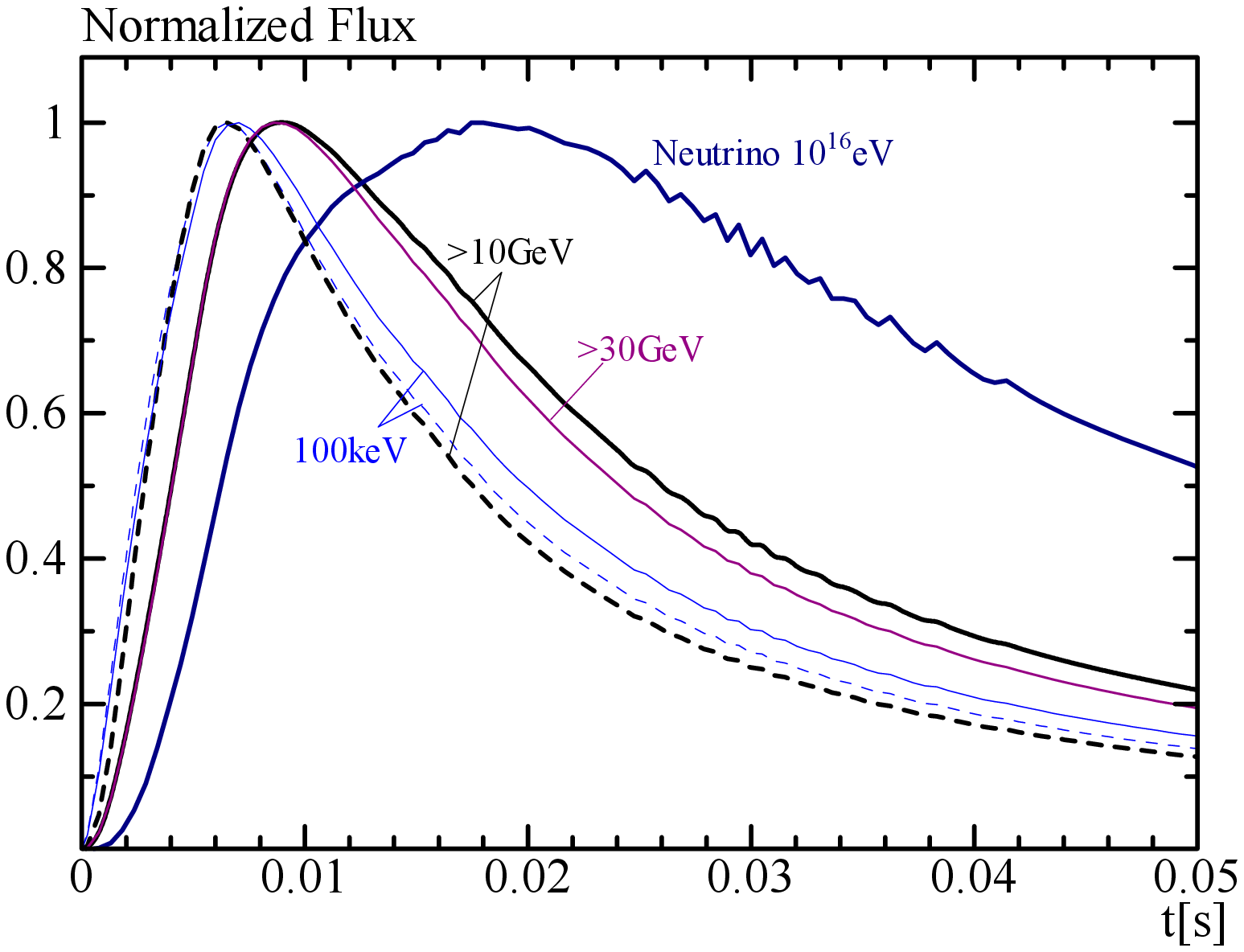}
\caption{Lightcurves for the ``moderate'' case with $z=1$.
See the text in \S \ref{sec:mod}.
The designation of the labels is the same as that in Figure \ref{fig:3}.
The solid lines correspond to the hadronic model,
while the dashed lines are for the same leptonic model without proton injection.
The neutrino flux evolution at $10^{16}$ eV is also plotted.
\label{fig:8}}
\end{figure}

\begin{figure}[htb!]
\centering
\epsscale{1.0}
\plotone{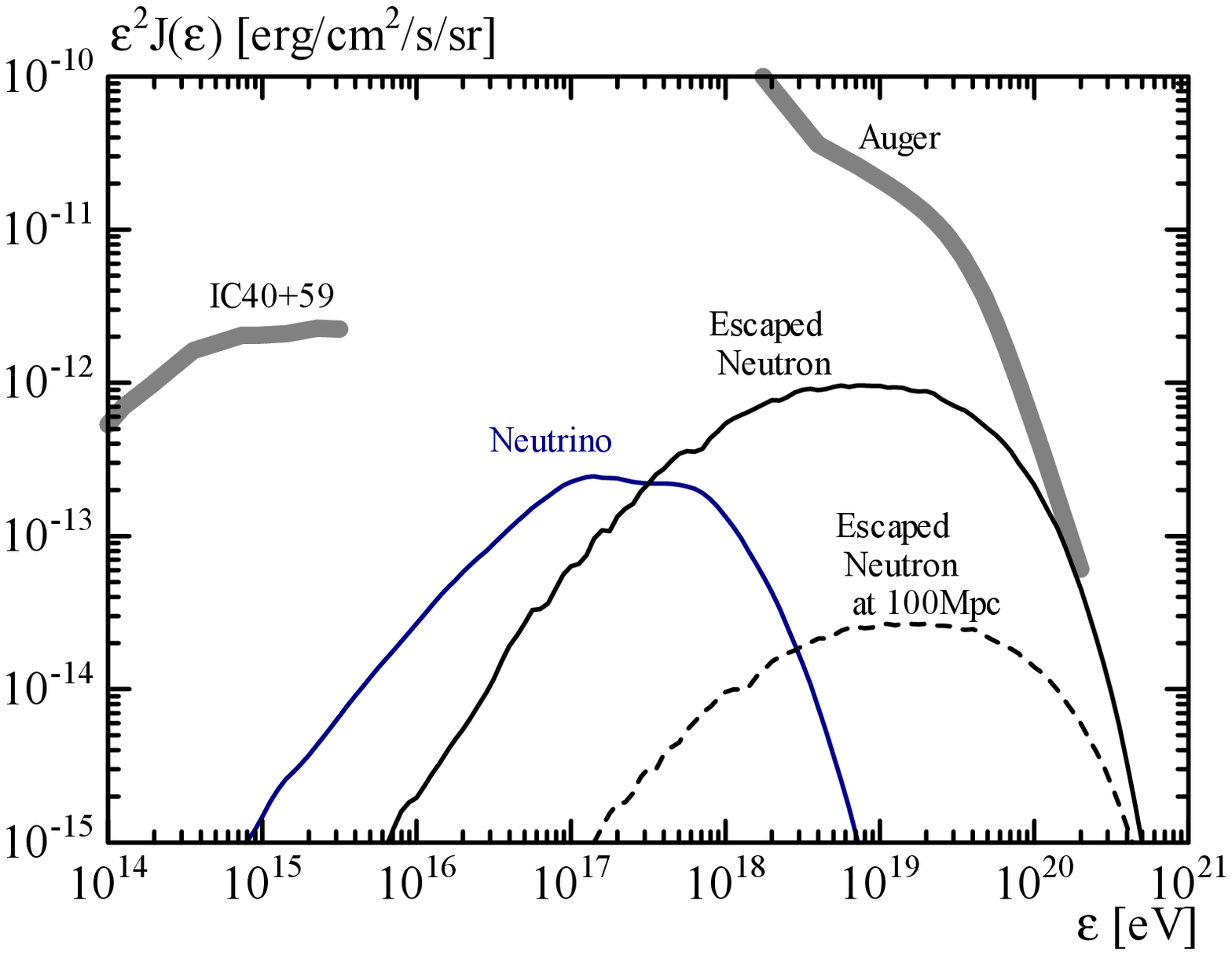}
\caption{Spectra of the diffuse neutrons and neutrinos
(without the GZK effect)
adopting a model corresponding to the ``moderate'' case ($\epsilon_{\rm p}/
\epsilon_{\rm e} \simeq $12), from $z=1$ (full lines,
accumulating 667 GRBs $\mbox{yr}^{-1}$),
and from 100Mpc (dashed line, for neutrons with a local rate
2 GRBs $\mbox{Gpc}^{-3}~\mbox{yr}^{-1}$); see the text in \S \ref{sec:mod}.
The upper limit for $\nu_\mu$ derived by IceCube \citep{abb12}
and the UHECR flux determined by the {\it Auger} team \citep{abr10}
are also plotted.
\label{fig:9}}
\end{figure}

In this example, the isotropic-equivalent energies of
escaped neutrinos and neutrons are $4.4 \times 10^{49}$ erg,
and $2.2 \times 10^{50}$ erg, respectively, per pulse.
For reference,
let us roughly estimate diffuse neutrino background and the contribution
of the escaped neutrons to the UHECR flux based on this specific parameter set.
Here we consider multiple pulses with the identical parameter set for a GRB.
The pulse number $N=30$ with the ``moderate'' parameter set yields
the total isotropic-equivalent
gamma-ray energy of $E_{\rm iso}=2.0 \times 10^{52}$ erg.
This energy scale can be used as a typical example.
The local GRB rate $\sim 1~\mbox{Gpc}^{-3}~\mbox{yr}^{-1}$
with the average energy $E_{\rm iso}=2.0 \times 10^{52}$ erg
corresponds to the gamma-ray energy release rate
of $2 \times 10^{43}~\mbox{erg}~\mbox{Mpc}^{-3}~\mbox{yr}^{-1}$.
The redshift $z=1$ in this example is lower than the observed
average redshift \citep{jak06}, so the total fluence
$7.4 \times 10^{-6}~\mbox{erg}~\mbox{cm}^{-2}$ per burst
is also slightly larger than the typical GRBs.
Simply accumulating this identical GRB at $z=1$
with $N=30$ and the GRB rate of $667~\mbox{yr}^{-1}$ \citep{abb12},
we plot the spectra of the neutrino background in Figure \ref{fig:9}.
Here we have not distinguished between the neutrino species, so
the spectrum consists of neutrinos originating from both pion-decay and muon-decay.
Even though the adopted gamma-ray fluence is larger than the average one,
the neutrino fluence is lower than the IceCube limit for $\nu_\mu$ \citep{abb12}.
This may be due to the large $R_0$ and $\Gamma$ in this model.
Furthermore, note that the dominant energy region for this model
is $10^{17}$--$10^{18}$ eV, which is well above the energy region 
constrained with IceCube.
Both in \citet{abb12} and here, the neutrino spectrum is consistent with the 
commonly considered case where neutrinos from pion-decay dominate, in which
the lower energy break of the neutrino spectrum \citep{wax97} is given by
\begin{eqnarray}
\varepsilon_{\nu,{\rm br}} &\simeq& \left( 1-\frac{m_\mu}{m_{\pi}} \right) K_{{\rm p}\gamma}
\gamma'_{\rm p,th} \delta \\
&\simeq& 8.6 \times 10^{16}
\left( \frac{1+z}{2} \right)^{-2}
\left( \frac{\varepsilon_{\rm peak}}{100 \mbox{keV}} \right)^{-1}
\left( \frac{\Gamma}{800} \right)^{2} \mbox{eV},
\label{eq:nubr}
\end{eqnarray}
where we take $\delta = 2 \Gamma$ as the Doppler factor, and $K_{{\rm p}\gamma}=0.2$.
As shown in \citet{asa09a}, a large Lorentz factor or a large radius $R_0$ 
is required in order to have GeV photons which escape unabsorbed from the source, 
and this leads to a lower photopion efficiency and a harder neutrino spectrum.
Therefore, such models which result in GeV emission can avoid the GRB-neutrino
upper limit constraints given by IceCube. Also, the report in \citet{abb12} 
excludes only GRB models with $\Gamma \lesssim 400$.

With the same method as for the neutrinos,
we also plot the spectrum of the escaped neutrons without
the GZK effect in Figure \ref{fig:9}.
The neutron energy range extends as far as the UHECR energy region,
and the contribution from those 667 GRBs in this simple model
seems close to the UHECR spectrum obtained with {\it Auger} \citep{abr10}.
The isotropic-equivalent energy of the neutrons escaped 
per burst (assuming $N=30$ pulses) is $6.6\times 10^{51}$ erg, comparable to the
energy emitted per burst (30 pulses) in $\gamma$-rays, $2\times 10^{52}$ erg.
However, at $10^{20}$ eV the GZK effect is crucial so that
we should count only GRBs within the GZK horizon ($\sim 100$ Mpc).
The GRB rate within 100 Mpc sphere ($\sim$ the GZK horizon)
becomes $8.4 \times 10^{-3}$ GRBs $\mbox{yr}^{-1}$
for the optimistic rate of $2~\mbox{Gpc}^{-3}~\mbox{yr}^{-1}$
\citep{wan10}.
Here we simply assume that GRBs occur with this rate at a distance of 100 Mpc,
and we plot the UHECR flux from those GRBs as
the dashed line in Figure \ref{fig:9}.
Note that even though we adopt $\epsilon_{\rm p}/\epsilon_{\rm e}\simeq 12$.
the energy fraction of escaped neutrons to gamma-rays is $\sim 1/3$, due to the
larger radius and $\Gamma$ leading to a modest $p\gamma$ rate.
The UHECR flux level at $10^{20}$ eV
is lower than the observed flux by a factor of $\sim 40$,
which implies that ten times the gamma-ray energy is required to
be emitted as cosmic rays to make GRBs dominant sources of UHECRs.
This is reconfirmation of the difficulty in GRB-UHECR scenario
pointed out in \citet{eic10}.
Here we have taken the most pessimistic assumption for UHECR production,
namely, that the high-energy protons cannot escape until they have cooled
adiabatically.
As a result, the escaped-neutron energy is a few percent of the injected energy 
of protons.  Some mechanism
to make protons escape may be required to produce UHECRs from GRBs.
Of course, we also have to include the luminosity function, redshift distribution,
and the uncertainties discussed in \citet{wax10}.
Such detailed estimate of neutrinos and UHECR with our numerical simulations
will be addressed in our future work.

\subsection{Hadronic ``Fermi-LAT'' Case}
\label{sec:lat}

Here we adopt a more specific parameter set that mimics a pulse of the
bright GRBs observed with {\it Fermi}-LAT.
The model parameters are $\Gamma=600$, $R_0=1.3 \times 10^{16}$ cm,
$E_{\rm e,pls}=2.0 \times 10^{54}$ erg,
and $E_{\rm p,pls}=4.8 \times 10^{55}$ erg
($\epsilon_{\rm p}/\epsilon_{\rm e} \simeq 24$) including the acceleration
effect after injection.
The initial magnetic field $B'_0=830$ G is determined by the same method
in \S \ref{sec:mod} as $\epsilon_B/\epsilon_{\rm e}=3.0$ at $R=2 R_0$.
The peak energy $\varepsilon_{\rm peak}$ is adjusted to be
$\sim 1$ MeV by adopting $\varepsilon'_{\rm e,min}=13$ GeV.

\begin{figure}[htb!]
\centering
\epsscale{1.0}
\plotone{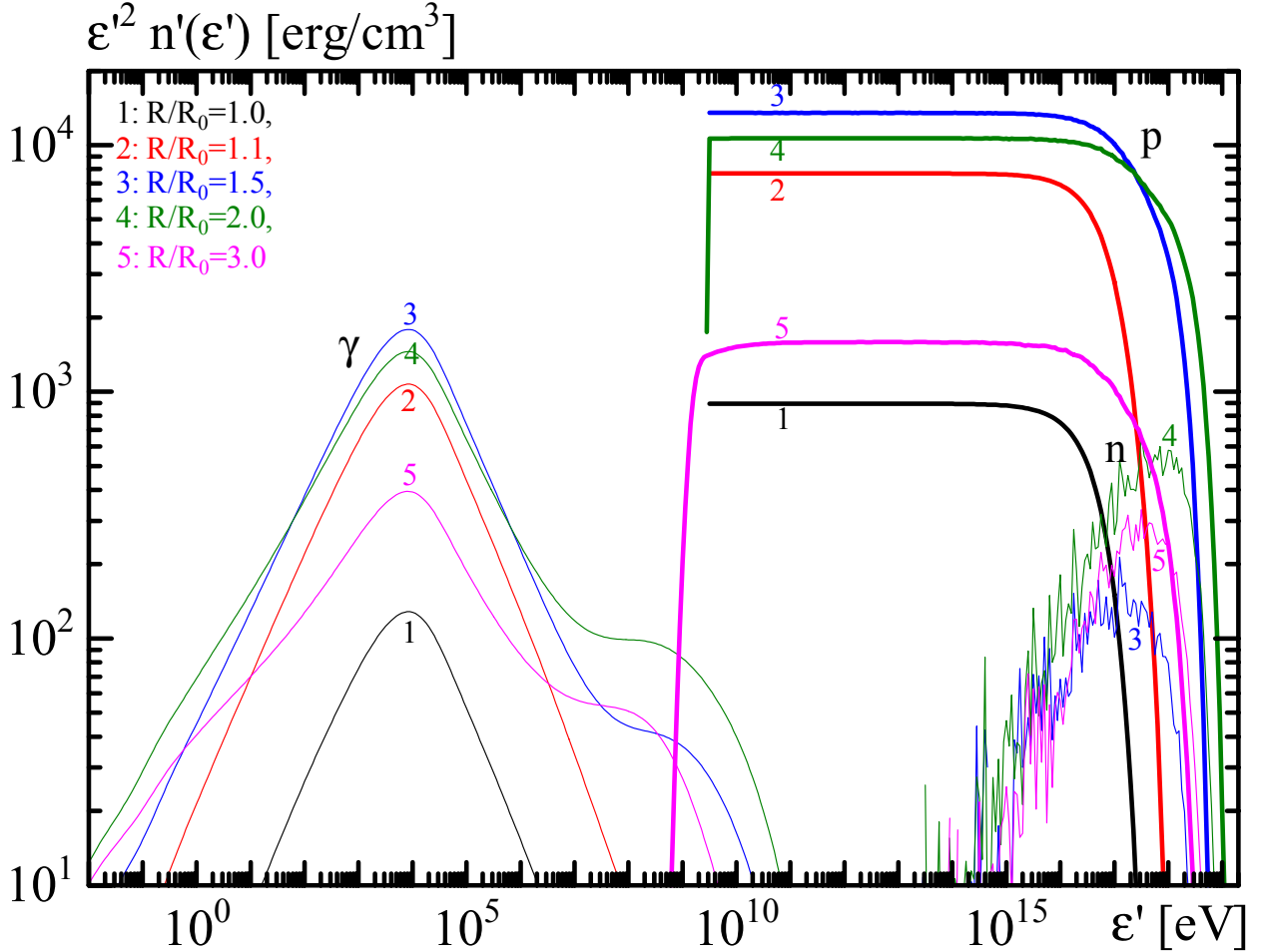}
\caption{Hadronic ``Fermi-LAT" case: spectral evolutions for photons, protons 
and neutrons in the shell frame with the expanding radius $R$
(see the text in \S \ref{sec:lat}).
The bump in the photon spectra at $10^7$--$10^{10}$ eV
is due purely to hadronic cascades.
\label{fig:10}}
\end{figure}

The evolution of the particle spectra in the shell frame
(Figure \ref{fig:10}) is similar to that for the ``moderate''
case in Figure \ref{fig:4}.
However, the gamma-ray spectral bump above $10^7$ eV 
is here attributed purely to hadrons.
Although the ratio $\epsilon_B/\epsilon_{\rm e}$ is the same as
that for the ``moderate'' case,
the Klein--Nishina effect is crucial
because of the very high $\varepsilon'_{\rm e,min}$.
Therefore, the IC emission does not yield a spectral bump in this case.
As the photon density increases, photons due to electromagnetic cascade
triggered by photopion production appear above $10^7$ eV.

\begin{figure}[htb!]
\centering
\epsscale{1.0}
\plotone{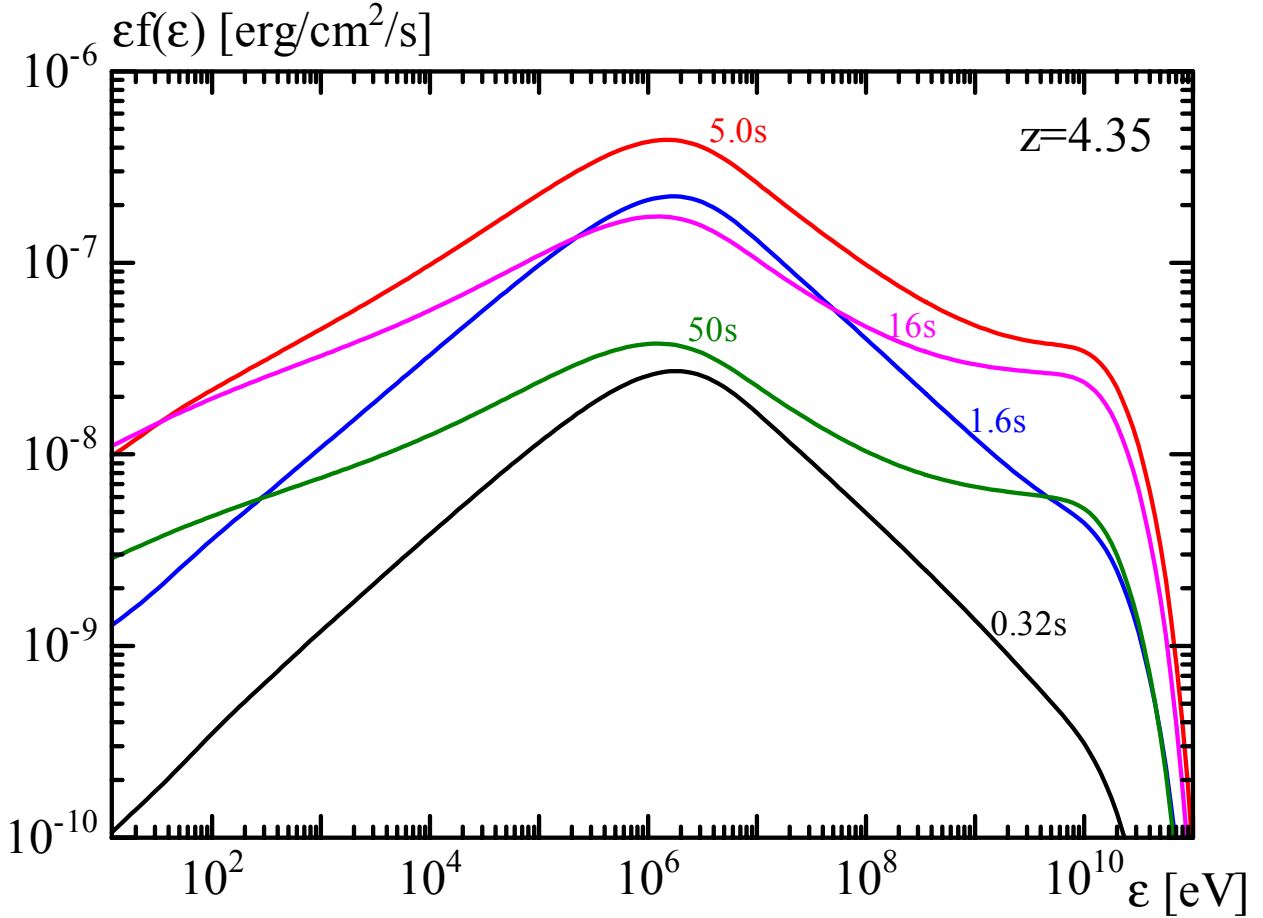}
\caption{Hadronic ``Fermi-LAT" case: flux evolution 
(see the text in \S \ref{sec:lat}) for an observer with $z=4.35$.
\label{fig:11}}
\end{figure}

\begin{figure}[htb!]
\centering
\epsscale{1.0}
\plotone{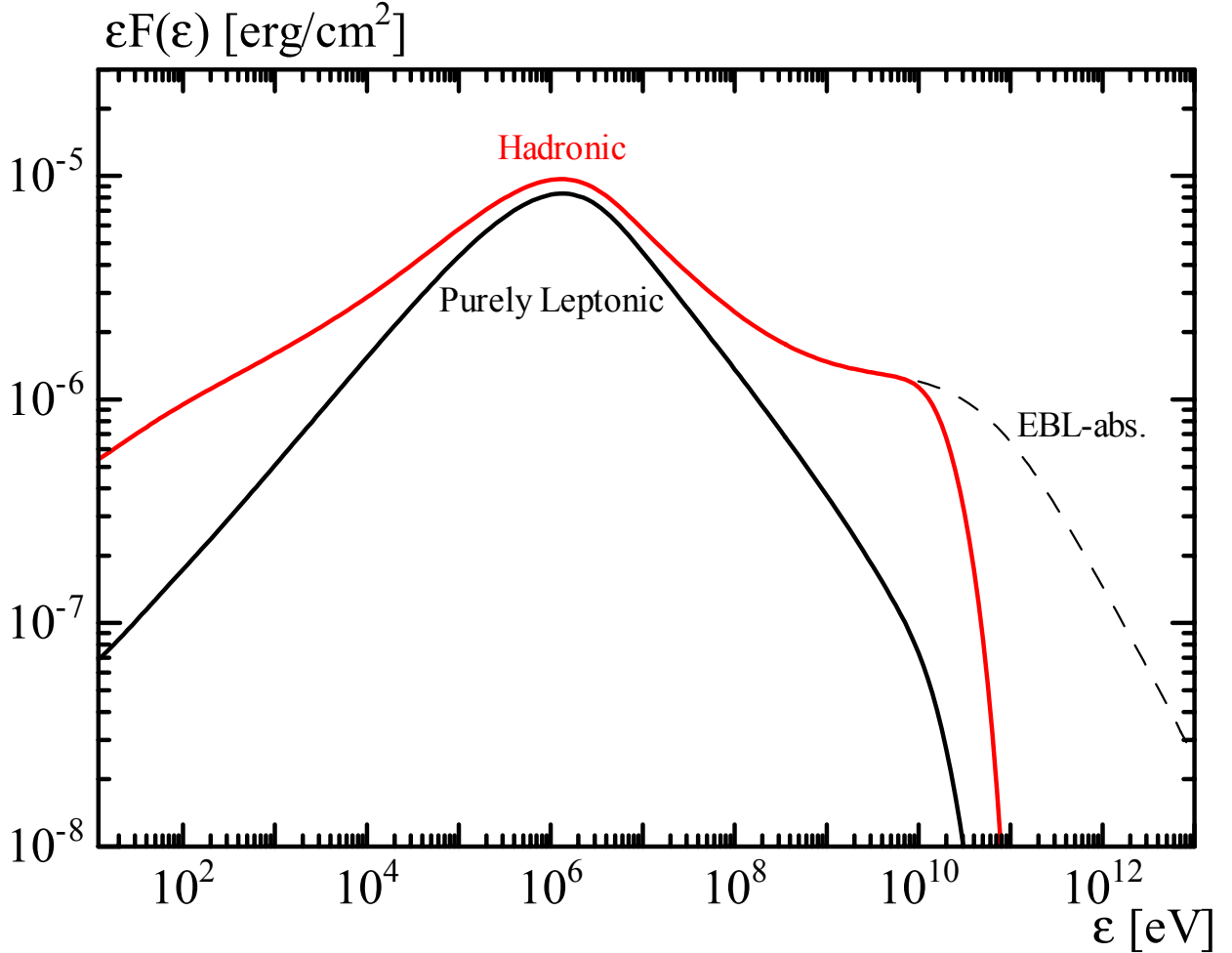}
\caption{Fluences for the hadronic model
and purely leptonic model ($\epsilon_{\rm p}=0$) with the ``Fermi-LAT'' parameters
(see the text in \S \ref{sec:lat}) for an observer with $z=4.35$.
The dashed line is the spectrum without the absorption effect of the EBL.
\label{fig:12}}
\end{figure}

Assuming $z=4.35$, we plot the observer-frame spectral evolution
in Figure \ref{fig:11}. Initially ($t \leq 1.6$ s) the spectral shape
is close to the simple Band function, and no spectral bump appears
in the GeV energy range.
In the later phase, the hadronic component appears, and
the entire shape gradually evolves into flat one.
This evolution is similar to the spectral behavior in GRB 080916C \citep{916C}.
The effect of the hadronic cascade on the spectral flatness
is well shown in the time-integrated spectra in Figure \ref{fig:12}.
In contrast to this two-component hadronic behavior, the purely leptonic case
for the same parameters leads to a pure Band-type time-integrated spectrum.
The spectral shape is almost unchanged during the emission
in the leptonic model.

\begin{figure}[htb!]
\centering
\epsscale{1.0}
\plotone{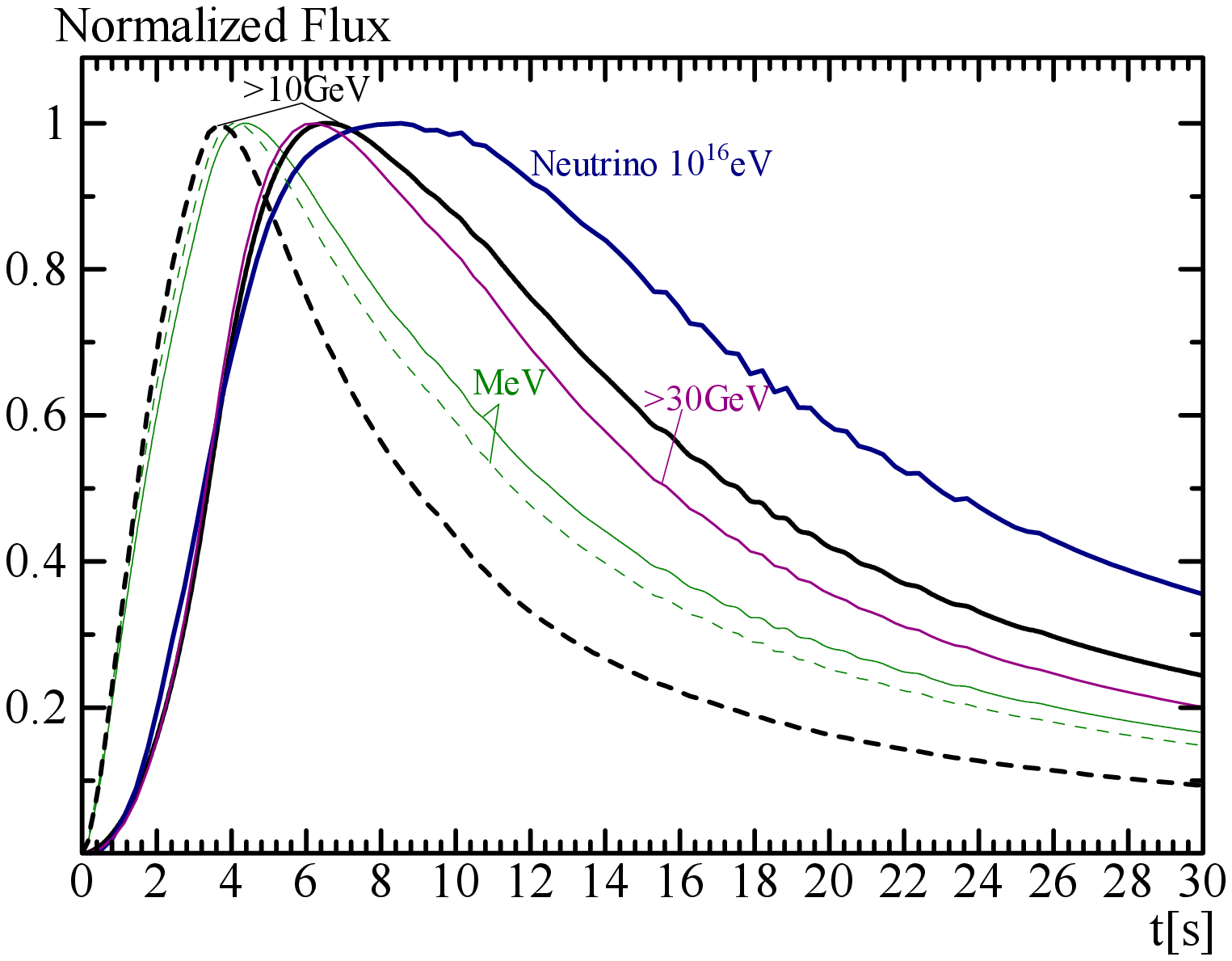}
\caption{Lightcurves for the ``Fermi-LAT'' case with $z=4.35$.
See the text in \S \ref{sec:lat}.
The designation of the labels is the same as that in Figure \ref{fig:8};
full lines are for the hadronic model, dashed are for the purely leptonic one.
In the hadronic model,
since the 10--30 GeV photons originate mainly from the hadronic cascades,
the peak times for those lightcurves and for the neutrino curve almost coincide
with each other.
\label{fig:13}}
\end{figure}

The lightcurve in Figure \ref{fig:13} shows a significant delay of
the 10 GeV-lightcurve.
The delay timescale ($\sim 6$ s) is comparable to
the delay seen in GRB 080916C \citep{916C}.
In this case SSC is negligible,
so the GeV spectral-bump is purely due to hadronic cascade.
Unlike in the ``moderate case'', the peak time of the 10 GeV photon
lightcurve is here relatively close to the peak time for the neutrinos.
Thus, the lack of SSC effects (moderated by the Klein--Nishina effect here)
is of key importance in emphasizing the delayed onset in hadronic models.
The 10 GeV-lightcurve shows a significantly broader FWHM ($\sim 14$ s)
than the FWHM of the MeV-lightcurve ($\sim 11$ s).
In the same model without proton injection (dashed lines),
the high-energy lightcurve is rather narrower than the MeV lightcurve.
The onsets of the GeV and MeV lightcurves are almost simultaneous.
Thus, for this parameter set, the difference between hadronic and leptonic models
is prominent.

\begin{figure}[htb!]
\centering
\epsscale{1.0}
\plotone{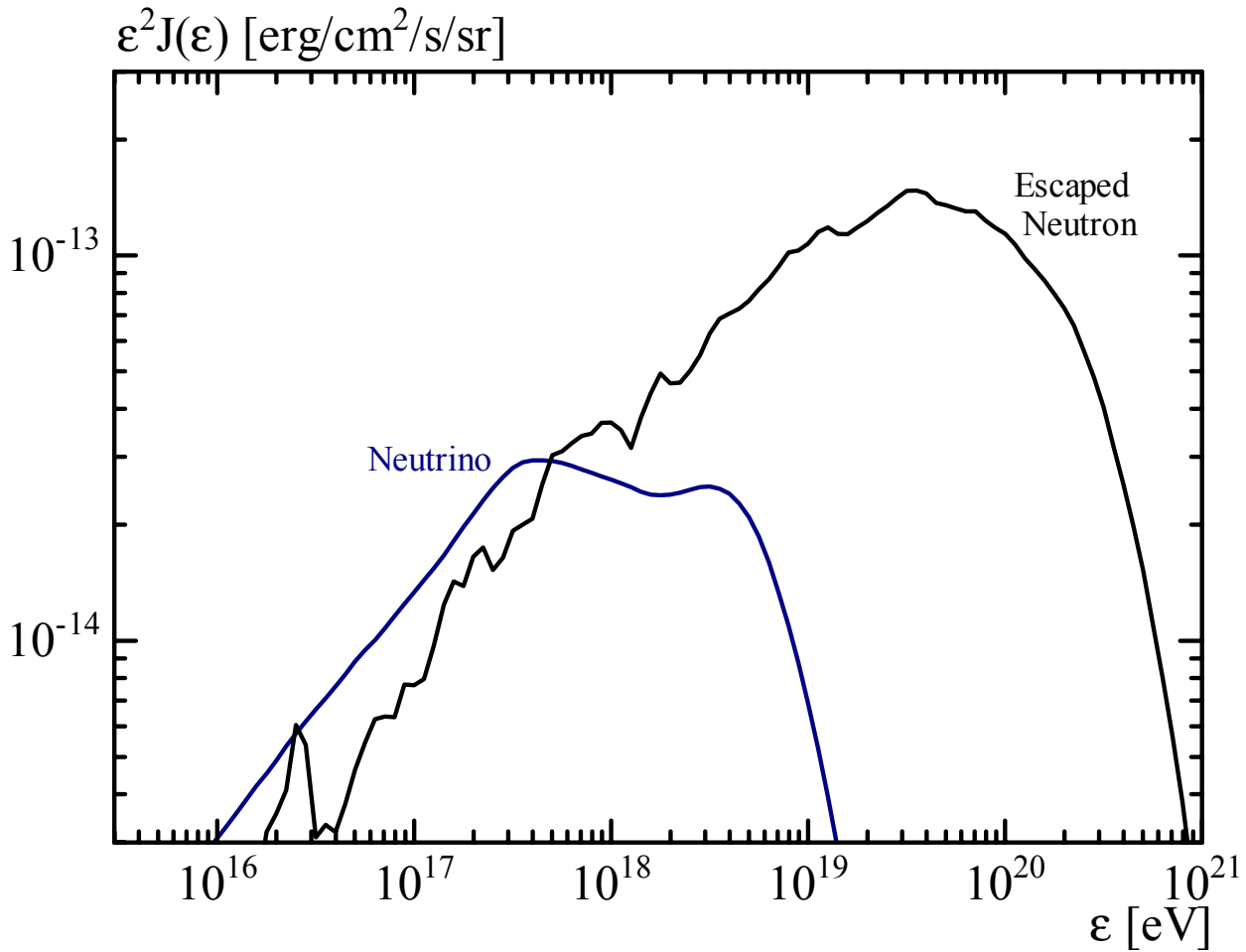}
\caption{Spectra of the diffuse neutrons and neutrinos
adopting a model with the ``Fermi-LAT'' hadronic cascade case
(accumulating 10 GRBs $\mbox{yr}^{-1}$).
See the text in \S \ref{sec:lat}.
\label{fig:14}}
\end{figure}

Here we have adopted a very large energy per pulse.
In such cases the number of pulses may be a few, and
the expected rate of such bright {\it Fermi}-LAT GRBs
(in this example the fluence is $7.5 \times 10^{-5}~\mbox{erg}~
\mbox{cm}^{-2}$) may be less than 10 per year.
Thus, in the same manner as in Figure \ref{fig:9}
adopting $N=1$ and the rate $10~\mbox{yr}^{-1}$,
we plot the spectra of the neutrino and neutron due to bright {\it Fermi}-LAT GRBs
in Figure \ref{fig:14}.
The spectrum of the escaped neutrons ranges as far as the UHECR energy scale
in this model also.
The contribution to the diffuse neutrino background
and to the integrated flux of escaping neutrons
are smaller than in the ``moderate case'', owing to the low GRB rate
with such parameters.
However, the spectral flux of the neutrons at $10^{20}$ eV is
comparable to that in the ``moderate case'',
although such neutrons lose their energy by the GZK process.
(Note that here we assumed $z=4.35$, and the equivalent rate
from within 100 Mpc would be much smaller than that for the 
``moderate" hadronic case parameters).
The isotropic-equivalent energy of neutrons per pulse
is $8.7 \times 10^{53}$ erg,
which is only $1.8$ \% of the injected energy of protons, while the 
emitted isotropic-equivalent gamma-ray energy per pulse
is $2.8 \times 10^{54}$ erg.

The isotropic-equivalent energy of the neutrinos per pulse
is $1.8 \times 10^{53}$ erg.
The break energy at $\sim 10^{17}$ eV in the neutrino spectrum
is in apparent contradiction
with equation (\ref{eq:nubr}) for this parameter set. The reason is that in
this case the pion production efficiency is not high enough,
even for protons of $\gamma'_{\rm p}=\gamma'_{\rm p,th}$,
the efficiency increasing as $\propto \gamma_{\rm p}^{0.5}$,
as mentioned in \S \ref{sec:model}.
The neutron spectrum in the ``Fermi LAT" case in Fig. \ref{fig:14} has a sharper 
peak at $\sim 3 \times 10^{19}$ eV than the ``moderate" model of Fig. \ref{fig:9}.
This indicates that only very high-energy protons near the cut-off energy
produce pions efficiently.
The high-energy bump at $\sim 3 \times 10^{18}$ eV in the neutrino spectrum
is attributed to pion-decays from such high-energy protons/neutrons,
while the low-energy break is attributed to the muon-decay component.
Thus, also for this parameter set, the neutrino energy range is well 
above the energy range constrained by IceCube.
The neutrino fluence ($\sim 10^{-6} ~\mbox{erg}~\mbox{cm}^{-2}$)
from this pulse corresponds to $\sim 1/300$ of the IceCube limit
\citep[$\sim 0.2 ~\mbox{GeV}~\mbox{cm}^{-2}$,][]{abb12}.

\subsection{Proton Synchrotron}
\label{sec:pro}

A different hadronic model is the proton synchrotron model \citep{vie95,tot98,raz10}.
To test this case with the same Lorentz factor and initial radius as those 
in \S \ref{sec:lat} we need to use extreme values for some of the other
parameters, in order to ensure that proton synchrotron contributes
to GeV energy range. Thus, we adopt
$E_{\rm e,pls}=4.0 \times 10^{54}$ erg,
and $E_{\rm p,pls}=2.0 \times 10^{56}$ erg
($\epsilon_{\rm p}/\epsilon_{\rm e}=50$),
and the initial magnetic field $B'_0=4800$ G
($\epsilon_B/\epsilon_{\rm e}=50$ at $R=2 R_0$).
The minimum energy of electrons is set as $\varepsilon'_{\rm e,min}=5.3$ GeV.
If the acceleration parameter $\xi$ for protons is unity,
the maximum energy of protons in this strong magnetic field
becomes too large and 
cascade emission from photopion production overwhelms
the proton synchrotron emission.
This is because the soft photon spectrum $\alpha \sim -1.5$
implies $f_{{\rm p}\gamma} \propto \gamma_{\rm p}^{0.5}$.
In order to suppress photopion production,
we degrade the maximum energy of protons by adopting $\xi=100$.

\begin{figure}[htb!]
\centering
\epsscale{1.0}
\plotone{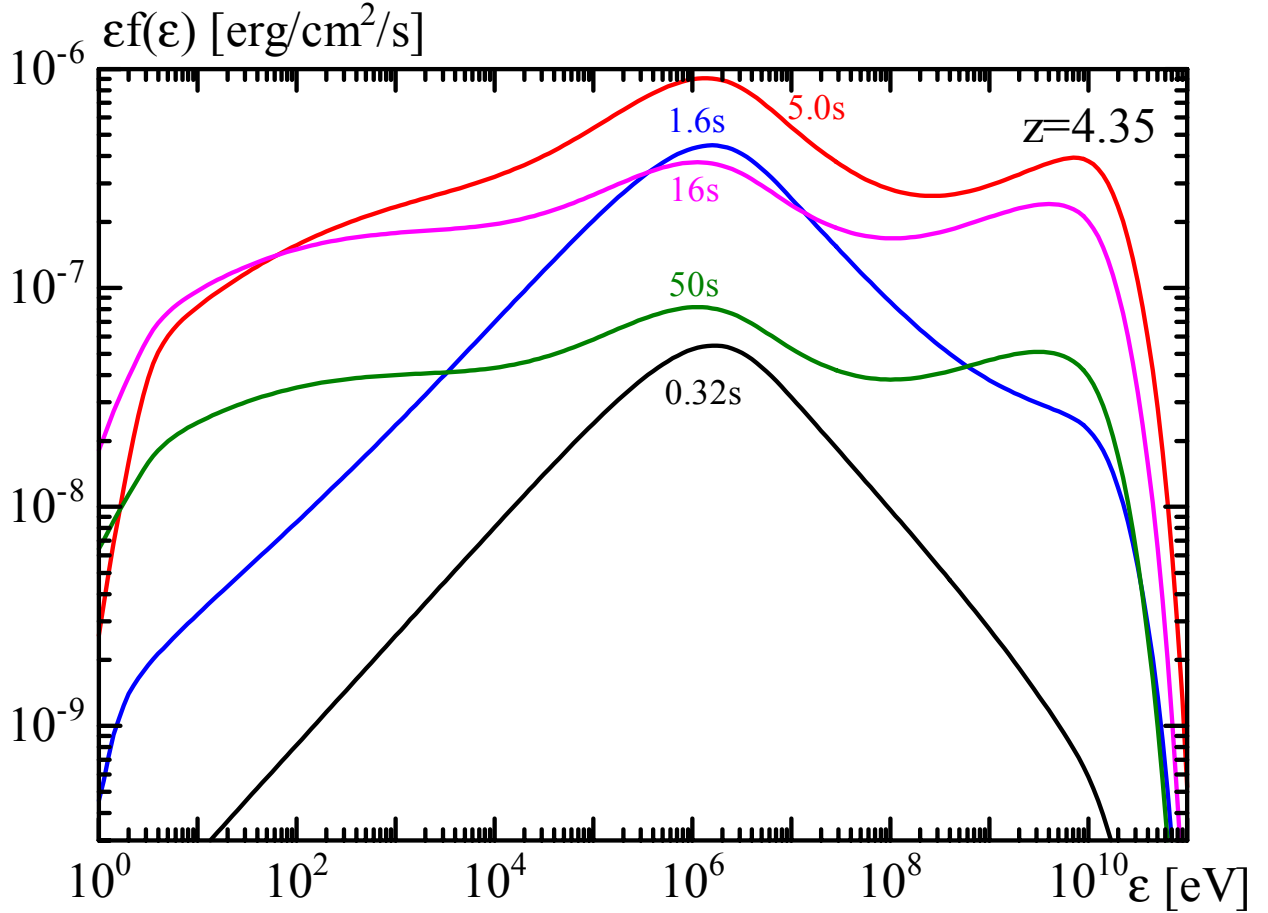}
\caption{Proton synchrotron model: flux evolution 
(see the text in \S \ref{sec:pro}) for an observer with $z=4.35$.
\label{fig:15}}
\end{figure}

\begin{figure}[htb!]
\centering
\epsscale{1.0}
\plotone{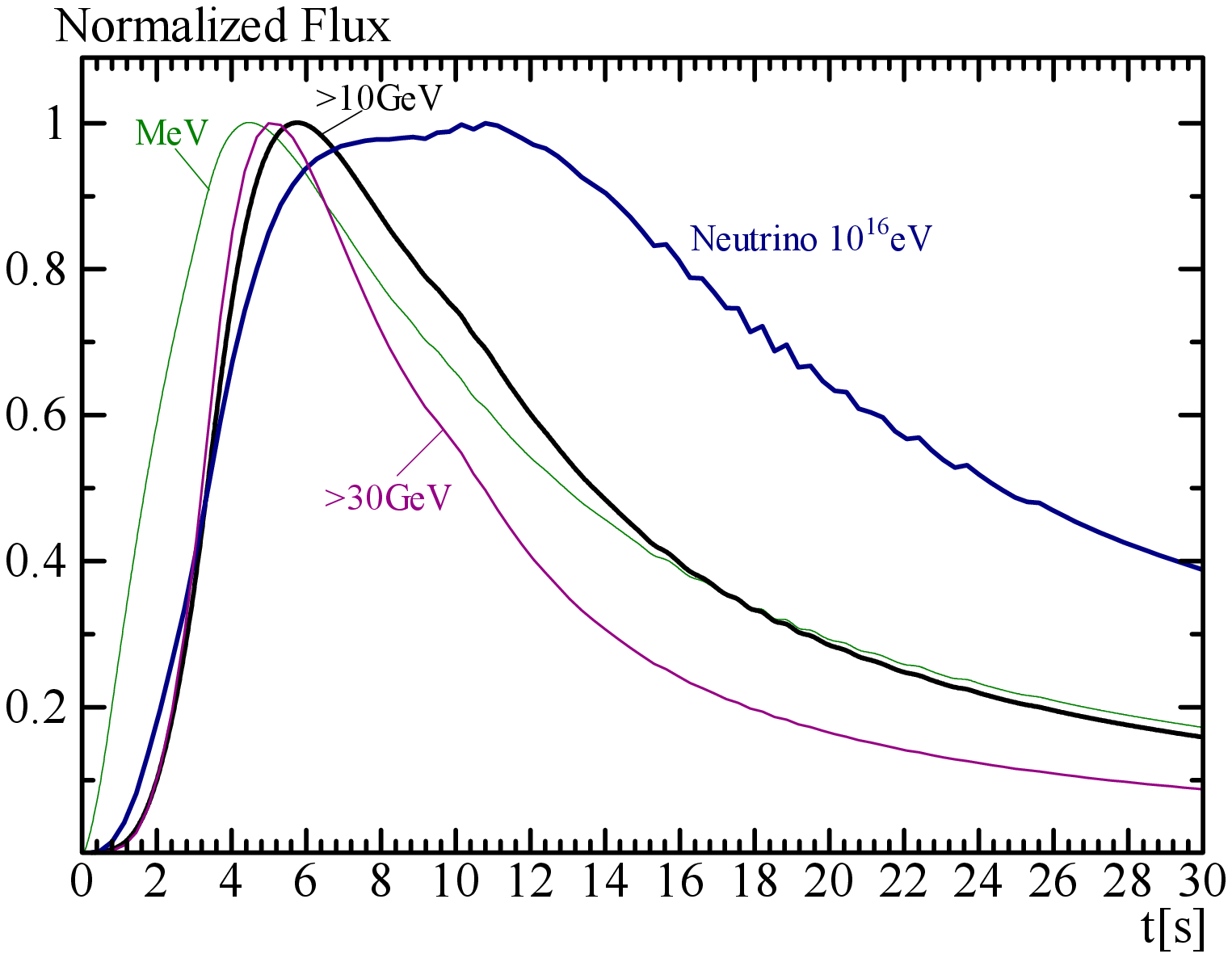}
\caption{Proton synchrotron model: lightcurves observed from $z=4.35$.
See the text in \S \ref{sec:pro}.
The designation of the labels is the same as that in Figure \ref{fig:8}.
\label{fig:16}}
\end{figure}

The resultant spectral evolution for an observer at $z=4.35$
is shown in Figure \ref{fig:15}.
Here the second peak of the photon spectra
is due to proton synchrotron radiation.
It is seen that proton synchrotron can produce harder spectra than
the cascade emission seen in Figure \ref{fig:11}.
Even in this case, however, we cannot neglect photopion production.
In the later phases, the spectrum becomes flat owing to the cascade
triggered by photopions.
The delay time of the 10 GeV-lightcurve (Figure \ref{fig:16}) is similar to
that for the ``Fermi-LAT'' case.
This is because the parameters $\Gamma$ and $R$ are common and
the GeV bump is due to purely hadronic effects in both models.
However, the FWHM of the 10 GeV-lightcurve is narrower here
and the decay timescale of the 10 GeV emission seems almost the same as
the timescale in the MeV emission, while the
30 GeV-lightcurve is narrower than the MeV lightcurve.
This depends on the redshift, or on model parameters ($\Gamma$ etc.).
However, for the proton synchrotron model,
the pulse decay timescale roughly corresponds to the synchrotron cooling timescale,
so that the decrease of the pulse timescale with photon energy
may be a characteristic feature of such models.
Although the degeneracy in the  parameters is not easy to resolve,
having an energy threshold as low as possible could help in distinguishing
the previous ``Fermi-LAT" hadronic cascade and this ``Fermi-LAT" proton 
synchrotron model through their lightcurve differences.

\begin{figure}[htb!]
\centering
\epsscale{1.0}
\plotone{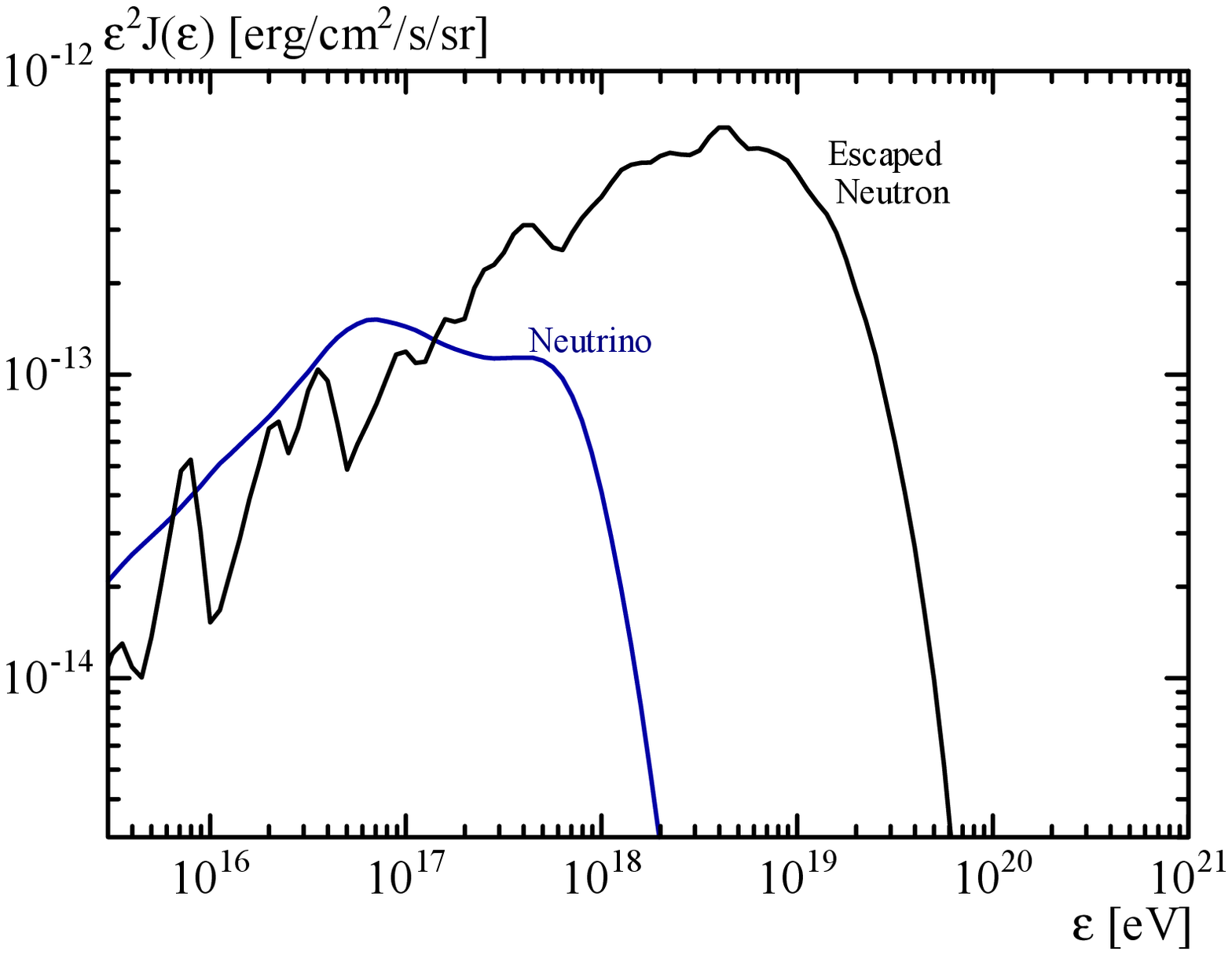}
\caption{
Spectra of the diffuse neutrons and neutrinos
adopting a model with the ``Fermi-LAT" proton 
synchrotron case.
See the text in \S \ref{sec:pro}.
\label{fig:17}}
\end{figure}

The emitted energy as gamma-rays, neutrons,
and neutrinos per pulse are $9.0 \times 10^{54}$ erg, $3.2 \times 10^{54}$ erg,
and $8.0 \times 10^{53}$ erg, respectively.
In Figure \ref{fig:17} we plot the spectra of the escaped neutrons
and neutrinos in the same manner ($N=1$, 10 GRBs $\mbox{yr}^{-1}$)
as in Figure \ref{fig:14}.
The maximum energy of neutrons is suppressed by the
large $\xi$ (see equation (\ref{acctim})). 
The neutron and neutrino spectral shape are similar to 
those in Figure \ref{fig:14}, respectively, although the typical 
energy ranges become lower.  This is because the critical parameters for 
photopion production ($R_0$, $\Gamma$, and $E_{\rm e,pls}$) are almost
the same in the two cases.

\section{Summary and Discussion}
\label{sec:sum}

We have simulated the time evolution of the particle energy distributions
in a relativistically expanding shell with parameters corresponding
to that of GRBs observed by {\it Fermi}, and discussed the spectral 
evolution of the radiation seen in the observer frame.
If we adopt a very weak magnetic field in the shell
in a one-zone leptonic model, the slow evolution
of the synchrotron self-Compton (SSC) emission
due to the Klein--Nishina effect results 
in a delayed onset of the GeV emission, by amounts comparable to those 
observed by {\it Fermi}.
However, in such leptonic models the FWHM of the GeV lightcurve is 
almost the same as that of the 0.1--1MeV lightcurves,
although the present-day low photon statistics in the GeV range
makes it hard to compare the FWHMs.
On the other hand, an  external inverse Compton (EIC) model, which is 
essentially a leptonic two-zone model, can reproduce the GeV-delay naturally.
A long tail for the GeV lightcurve, as shown in AM11,
is possible evidence for this model.
This can partially contribute to the extended GeV emission
seen in several GRBs \citep[see, e.g.,][]{ghi10}.
We have also tested the $\gamma \gamma$-opacity evolution effect with our code.
At least in our one-zone formulation, an electron injection timescale longer
than the expansion timescale is required in order to observe the opacity damping.
The resultant lightcurves have a characteristic shape, and the growth of the 
cutoff energy is more gradual than in the current sample of 
{\it Fermi}-LAT bursts, which so far do not show a similar spectral evolution.
Thus, $\gamma \gamma$-opacity evolution cannot be the reason for the delay.

The hadronic models studied here include both cascade models and a  proton 
synchrotron model, both of which are able to reproduce the delayed onset
with a delay timescale close to the pulse timescale ($\sim R_0/c \Gamma^2$).
The delay is due to the long acceleration timescale of protons
and continuous photopion production after the end of the particle injection.
The wider FWHM for the GeV lightcurve than for then MeV lightcurve
could be a signature of the hadronic cascade.
If the Klein--Nishina effect prevents IC emissions,
the delay due to hadronic cascade becomes more dominant.
The amounts of escaped UHE neutrons in our examples
are comparable to the gamma-ray energy.
Since we have adopted large $\Gamma$ and $R$ values in order to 
simulate GRBs where GeV photons are able escape from the source, 
the resultant neutrino spectra are hard enough to avoid the 
current flux limit constraints from IceCube.
In addition, as the {\it Fermi} team indicates \citep{ack12}, GRBs 
with extra components in the GeV band may be a small fraction of the GRB population.
Therefore, even if GRBs accelerate UHE particles, the typical GRB parameters
are in a range which implies a sufficiently low neutrino and GeV gamma-ray
production efficiency.
Our hadronic model also predicts a delayed onset of the neutrino emission, which 
is more pronounced than the corresponding GeV photon delay. If neutrinos are
eventually observed, this may be another point which could be tested by 
future neutrino observations.
The energy budget required for being UHECR sources appears insufficient in 
our examples.  In future work, we plan to include the redshift evolution, 
luminosity function, and other details to further test the viability of the 
GRB-UHECR scenario.

The open problem of the low-energy photon spectral index $\alpha$ is not 
addressed in this paper. Here we inject the electrons in the usual manner;
the  minimum energy of the electrons at injection $\varepsilon'_{\rm e,min}$
being a free parameter\footnote{As is well known,
$\varepsilon'_{\rm e,min}$ is written by the phenomenological
parameters: $\epsilon_{\rm e}$ and the number fraction of the accelerated
electrons, both of which are not well constrained so far.}.
The synchrotron cooling results in a soft effective 
electron distribution ($n'_{\rm e}(\varepsilon'_{\rm e})
\propto \varepsilon_{\rm e}^{\prime -2}$), so $\alpha \lesssim -1.5$
is unavoidable in this type of simulations. Spectral slopes such as these, 
while softer than the average $\alpha \sim -1$, are present in the BATSE data base.
Such softer slopes have the result of enhancing the photopion production 
efficiency, as well as the resultant flux of escaped neutrons.
Note that here the energy fraction in protons required to generate a GeV 
extra component is not so high, $\epsilon_{\rm p}/\epsilon_{\rm e} =10$--$25$
($\sim f_e^{-1}$, in alternate notation), compared with the case of GRB 090510 where 
$\epsilon_{\rm p}/\epsilon_{\rm e} >100$ 
was required to reconcile with the observed index $\alpha=-0.58$ \citep{asa09a}.
Another consequence is that the soft spectra make it difficult to find a 
low-energy spectral excess in the X-ray region.
In our hadronic examples, the low-energy excess appears
in the later stages, when the power-law component due to the
hadronic cascade becomes prominent or dominant over the entire spectrum.
On the other hand, in order to have a self-consistent one-zone synchrotron model
whether leptonic or hadronic,
some mechanism to harden the spectrum 
\citep[e.g.,][]{asa09c,mur12} would need to be included, which could change the 
lightcurves or the UHECR production efficiency.

In our simulations the minimum-Lorentz factor of electrons
at injection $\gamma'_{\rm e,min}$ is assumed, for simplicity, to be constant
during the injection.  Thus, our simulations do not show a significant 
$\varepsilon_{\rm peak}$-evolution, as seen in Figure \ref{fig:11}, whereas
an increase of $\varepsilon_{\rm peak}$ around the time of the GeV onset
has been reported in GRB 080916C and 090902B.  To reproduce such a 
$\varepsilon_{\rm peak}$-evolution in our model, a growth of $\gamma'_{\rm e,min}$ 
during the injection would need to be introduced.  However, in GRB 090926A the 
$\varepsilon_{\rm peak}$ does not change drastically around the GeV onset time 
\citep{926A,zha11}.  Given this variety of spectral evolution behaviors, at this
stage one is unfortunately left without a definite guideline for including a temporal 
evolution of the model parameters.

In summary the GeV delays can be explained either by a one-zone leptonic 
model with very low magnetic field and high luminosity (AM11),
or by normal parameter range one-zone 
hadronic models (this work), or also by a normal magnetic field and 
luminosity two-zone leptonic model via EIC \citep{tom09,tom10}.
However, the lightcurve behavior of these models differ substantially from each other.
The qualitative tendency of the multi-GeV lightcurves discussed here
can be tested with future atmospheric Cherenkov telescopes such as CTA.
For this purpose, it would be desirable that the low-energy threshold
of those telescopes should be as low as possible, in order to ensure 
good photon statistics as well as the capability to observe the spectral 
evolution below the $\gamma \gamma$ cutoff energy due to EBL. Such an 
ideal instrument would provide the capabilities for significant breakthroughs 
in GRB physics.

\begin{acknowledgments}
First we appreciate the anonymous referee for valuable comments.
This study is partially supported by Grants-in-Aid for Scientific Research
No.22740117 from the Ministry of Education,
Culture, Sports, Science and Technology (MEXT) of Japan,
and NSF PHY-0757155.
We also thank the CTA consortium members, especially S. Inoue, K. Ioka,
and M. Teshima.
\end{acknowledgments}

\end{document}